\documentclass[
tightenlines,
superscriptaddress,
 preprint,
 amsmath,amssymb,
 aps,
]{revtex4-1}

\usepackage{graphicx}
\usepackage{bm}
\usepackage{amsmath}
\usepackage{xcolor}
\usepackage{braket}

\begin{document}

\title{Corner and side localization of electrons in irregular hexagonal semiconductor shells}

\author{Anna Sitek}
\affiliation{School of Science and Engineering, Reykjavik University, 
         Menntavegur 1, IS-101 Reykjavik, Iceland}
\affiliation{Department of Theoretical Physics, 
         Faculty of Fundamental Problems of Technology,
         Wroclaw University of Science and Technology, 
         50-370 Wroclaw, Poland}  
         
\author{Miguel Urbaneja Torres}
\affiliation{School of Science and Engineering, Reykjavik University, 
         Menntavegur 1, IS-101 Reykjavik, Iceland}
         
\author{Andrei Manolescu}
\affiliation{School of Science and Engineering, Reykjavik University, 
         Menntavegur 1, IS-101 Reykjavik, Iceland}

\begin{abstract}
We discuss the low energy electronic states in hexagonal rings. These
states correspond to the transverse modes in core-shell nanowires built
of III-V semiconductors which typically have a hexagonal cross section.
In the case of symmetric structures the 12 lowest states (including
the spin) are localized in the corners, while the next following 12
states are localized mostly on the sides.  Depending on the material parameters,
in particular the effective mass, the ring diameter and width, the
corner and side states may be separated by a considerable energy gap,
ranging from few to tens of meV.  In a realistic fabrication process
geometric asymmetries are unavoidable, and therefore the particles are not
symmetrically distributed between all corner and side areas.  Possibly,
even small deformations may shift the localization of the ground state to
one of the sides.  The transverse states or the transitions between them
may be important in transport or optical experiments.  Still, up to date, 
there are only very few experimental investigations of the 
localization-dependent properties of core-shell nanowires.
\end{abstract}

\maketitle

\section{Introduction}

Core-shell nanowires are radial heterojunctions consisting of 
a single-material nanowire (core) which is covered with one or more 
layers of different material (shells). Due to the crystallographic 
structure the cross section of such wires is usually 
hexagonal \cite{Blomers13,Rieger12,Haas13,Funk13,Erhard15,Weiss14b,Jadczak14}, 
but other shapes like triangles \cite{Qian04,Qian05,Baird09,Heurlin15,Dong09,Yuan15,Goransson19} 
or dodecagons \cite{Rieger15} have already been obtained. These radial 
heterojunctions have been in the focus of extensive experimental 
\cite{Jadczak14,Fickenscher13,Funk13,Shi15,Erhard15,Weiss14b,Kinzel16,Li17v2,DeLuca13,Plochocka13}
and theoretical \cite{Buscemi15,Ferrari09b,Wong11} studies in the recent 
years. 
This is mostly due to the possibility of controlling some of their physical properties,
e.g., the band alignment. 
If the materials are properly adjusted, and also the geometric variables such as
core diameter 
and shell thickness, then one may obtain type II band alignment at the 
heterojunction because of which the electrons are confined only in the 
shell volume and form conductive shells \cite{Blomers13,Li18}. Moreover, 
the core part may be etched out, and thus hollow systems or prismatic 
nanotubes of finite thickness may be obtained \cite{Rieger12,Haas13}.

Ever since the paper by Ferrari \textit{et al.} \cite{Ferrari09b}, it
has been known that low-energy electrons confined in thin hexagonal
tubes are accumulated along the edges, while the particles excited
to higher energies occupy the facets.  More generally, the shape
of the cross section governs the energy structure of prismatic
nanotubes \cite{Sitek15,Sitek16,Sitek18,Urbaneja18}. If the cross section
is a regular polygon with $N$ corners, and the thickness of the wall is
much smaller than the radius of the tube, the lowest $2N$ energy states
(including spin) are localized in the corners of the polygon.  Moreover,
the wave functions of the next group of $2N$ states, on the energy scale, are 
localized on the
sides of the polygon and are separated from the corner states by
an energy gap which, depending on the geometry and material parameters,
may be comparable or even larger than the room-temperature energy, especially
for the triangular case \cite{Sitek15,Sitek16}.  

For regular polygons, both the corner and side states have an internal
energy dispersion.  Still, this energy dispersion decreases with the width, 
eventually reducing to a quasidegenerate group of corner states.  On the contrary,
the corner localization softens with
increasing the aspect ratio of the polygon, i.e., the ratio between side thickness
to polygon diameter.  The probability maxima of the lowest energy states,
and in particular of the ground state, remain centered along the edges,
but spread and penetrate into the facets where they overlap. As a result,
the electrons are distributed around the entire circumference of the cross
section.

The presence of multiple corner and side states separated by a large
energy gap implies very interesting physics, like the absorption
of photons from different spectral domains \cite{Sitek15,Sitek16},
multiple Majorana modes in interaction \cite{Manolescu17,Stanescu18},
irregular conductance steps in ballistic transport \cite{Urbaneja18},
or spin-singlet pairs of Coulomb coupled electrons with energies within
the gap \cite{Sitek17,Sitek18}. These mentioned phenomena are only predicted
and they need the experimental detection.  
In reality, most of the reported experimental studies do not resolve the shape of the
wire and the results obtained from hexagonal structures can be
well approximated with a cylindrical model even for narrow shells
\cite{Fickenscher13,Gul14,Rosdahl14,Heedt16}.  Only in rare situations
the anisotropic electron distribution has been detected \cite{Funk13}.
As a result, not much attention has been paid to the shape or to the
internal structure of the wires and to the effects resulting from the 
inhomogeneous electron distribution within the cross section.  Very recently, 
two experimental papers have reported complex photoluminescence spectra
associated to irregular core-shell nanowires.  One of them \cite{Battiato19}
describes polychromatic emission of InP-InAs-InP multi-shell nanowires with 
a diameter that increases along the nanowire axis. The other one \cite{Sonner19}
shows spectra that could be attributed to exciton recombinations on different
facets situated on different regions along the length of an irregular GaAs shell
embedded in a GaAs-AlGaAs multiple heterojunction. 

Motivated by these recent experimental achievements, in this paper we compare the localization 
patterns of electrons on symmetric and irregular hexagonal rings. We consider several 
material parameters and show how the separation between corner and side states increases 
with decreasing the effective mass and how the corner and side localization 
and their separation evolve when the hexagonal symmetry is broken.
The paper is organized as follows. In the following section we present the model and 
the calculation method. Then, in Sec.\ \ref{Sym} we consider single 
(Sec.\ \ref{Sym_single}) and many-body (Sec.\ \ref{Sym_many}) energy levels and the
localization of symmetric quantum rings. Next, in Sec.\ \ref{Def} we study asymmetrically
(Sec.\ \ref{Def_asym}) and symmetrically (Sec.\ \ref{Def_sym}) deformed structures. 
Finally, in Sec.\ \ref{Con} we summarize the results.

\section{Model}

\begin{figure}[h]
	\centering
	\includegraphics[scale=0.10]{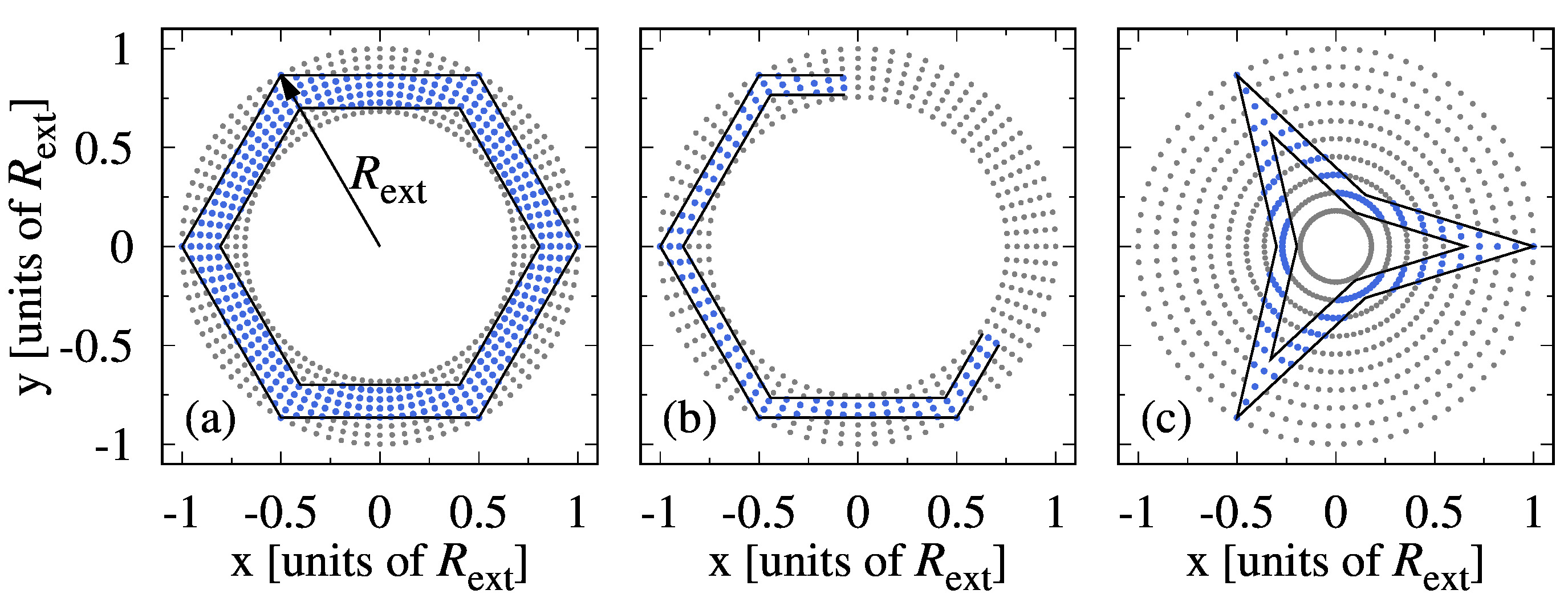} 
	\caption{Sample models: Polygonal constraints (black solid lines) 
			applied on a polar grid (grey and blue circles) which is further 
			reduced to the points situated between the boundaries (blue circles).	 
			The black arrow in Fig. (a) indicates the external radius of the 
			polar grid and of the polygons ($R_{\mathrm{ext}}$). 
			For visibility, we considerably reduced the number of side points.}              	               
	\label{fig_samples}
\end{figure}

If the shells are so short that the separation between the two lowest longitudinal 
modes is larger than the energy dispersion of the interesting range of transversal 
states then the shells may be considered quantum rings. In our case it is enough 
if the separation exceeds the gap between corner and side states and their energy 
dispersion.  

We model a polygonal quantum ring starting from a circular disk situated in
the $(x,y)$ plane. The plane is then is discretized on a polar grid \cite{Daday11},
on top of which we impose lateral boundaries conditions, corresponding to infinite
potential barriers, that enclose the hexagonal rings. The points that lie outside the defined boundaries are excluded, Fig.\ \ref{fig_samples}. The flexibility of this method allows us to model asymmetric samples without the necessity to adapt the background grid or redefining the Hamiltonian matrix elements. 

The Hilbert space associated with the polar lattice is spanned by the vectors $\vert q\rangle = \vert k_qj_q \rangle = \vert kj \rangle$ , in which the radial ($k$) and the angular ($j$) coordinates are included. The single-particle eigenvalues $E_{a}$ ($a=1,2,3, \ldots$) and the eigenvectors $\psi_{a} = \sum_{q}\psi(q,a)\ket{q}$ in the position representation are obtained through numerical diagonalization \cite{Sitek15} of the 
the Hamiltonian in the basis including the spin ($\sigma = \pm 1$), i.e., $\vert kj\sigma\rangle$. 

In the absence of an external electromagnetic field the system evolution is governed 
by the Schr\"odinger equation
\begin{equation}
\frac{p_z^2}{2 m_{\rm eff}}\psi_a = E_a \psi_a \ .
\label{single}
\end{equation}
The corresponding Hamiltonian matrix elements are 
\begin{eqnarray*}
\label{Hamiltonian}
\langle kj\sigma\vert \frac{p_z^2}{2 m_{\rm eff}}\vert k'j'\sigma'\rangle 
&=&
T\delta_{\sigma,\sigma'}\left[ t_r \left(\delta_{k,k'}
-\delta_{k,k'+1}\right)\delta_{j,j'}\right. \\
&+& \left. t_{\phi}\delta_{k,k'}\left(\delta_{j,j'}-\delta_{j,j'+1}\right) 
+ \mathrm{H.c.}\right], 
\end{eqnarray*}
where $T=\hbar^{2}/(2m_{\mathrm{eff}}R^{2}_{\mathrm{ext}})$ 
is a reference energy and $m_{\mathrm{eff}}$ is the effective mass of the electrons in the ring material, with $R_{\mathrm{ext}}$ the external radius of the polygonal shell. The factors $t_r=(R_{\mathrm{ext}}/\delta r)^2$ and 
$t_{\phi}=[R_{\mathrm{ext}}/(r_k\delta\phi)]^2$. Finally, $\delta r$ and $\delta\phi$  are the distance 
between neighboring sites with the same angle and the angle  difference between adjacent sites with the same radius, respectively.

The next step in the calculations is to use a subset of the single-particle eigenvectors $\psi_a$ as a 
basis for the many-body problem. The electron-electron interaction within the polygonal ring is considered 
by taking the Schr\"odinger equation in the second quantization formalism. The many body Hamiltonian is 
then solved again by numerical diagonalization
\begin{equation}
\hat H = \sum_a E_a c^{\dagger}_a c_a 
+ \frac {1}{2} \sum_{a,b,c,d} V_{abcd} c^{\dagger}_a c^{\dagger}_b c_d c_c \ ,
\label{many}
\end{equation}
where $c^{\dagger}$ and $c$ are the well known creation and annihilation operators, and 
the matrix elements of the Coulomb potential are calculated as
\begin{equation}
V_{abcd} = \langle \psi_a\psi_b | \frac{e^2}{4\pi\epsilon\epsilon_{0} |\bf r - \bf r'|}|\psi_c\psi_d\rangle \ ,
\label{Vcoul}
\end{equation}
where ${\bf r}=(x,y)$ is the position of electrons in the plane and
$\epsilon$ represents the relative dielectric permittivity of the shell
material, Table\ \ref{tab}.  The Hamiltonian (\ref{many}) is diagonalized
using a subspace of up to 24 corner and side states with the lowest energy
(configuration-interaction method) \cite{Sitek17}.


\begin{table}[h]
\caption{\label{tab} Material parameters used in the numerical simulations}
\footnotesize
\begin{tabular}{@{}cccccc}
\toprule
                         &            & InP         & GaAs         & InAs         & InSb \\
 effective mass          & $m_{\mathrm{eff}}$    & $0.08m_{e}$ & $0.067m_{e}$ & $0.023m_{e}$ & $0.014m_{e}$ \\
 relative permittivity   & $\epsilon$ & $12.5$      & $12.9$       & $15.0$       & $16.8$ \\
\toprule

$m_{e}$ is the free electron mass
\end{tabular}\\

\end{table}
\normalsize


\section{\label{Sym} Symmetric rings/shells}

\subsection{\label{Sym_single} Single-particle energy levels and localization}

\begin{figure}[h]
\centering
 \includegraphics[scale=0.095]{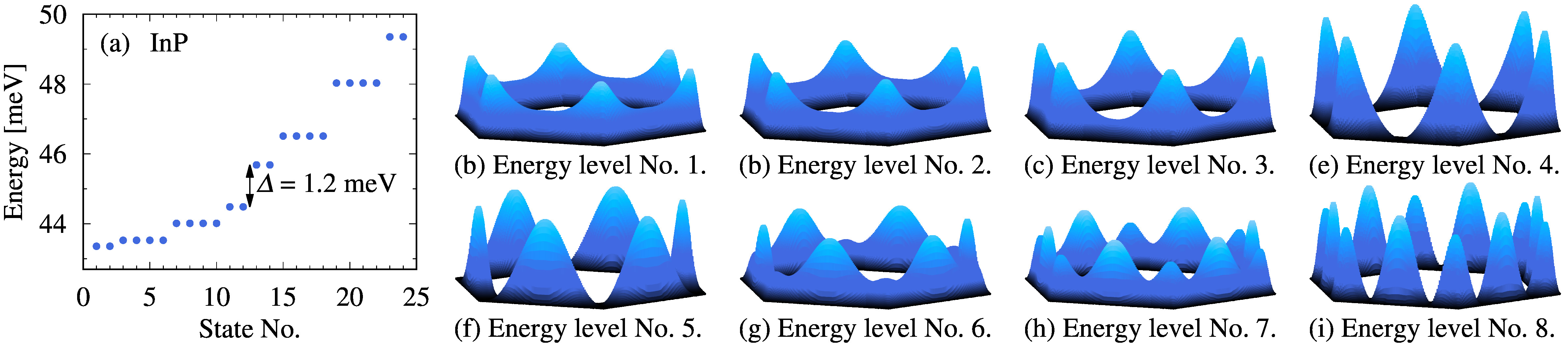} 
\caption{(a) Energy levels of a single electron confined in a symmetric InP ring 
         for which the side thickness and external radius are equal to 10 and 60 nm, 
         respectively. $\Delta$ stands for the gap between corner and side states.
         (b-i) The corresponding probability distributions in increasing energy order.     
         }   
\label{fig_sym_10nm}
\end{figure}

The ground state of a single electron confined in a hexagonal ring is 
spin degenerate, and it is followed by alternating pairs of four- and twofold
degenerate levels. The degeneracy is either due to spin and orbital momentum, or due to spin 
only, such that every four consecutive levels (12 states) repeat the degeneracy 
pattern 2-4-4-2.
In Fig.\ \ref{fig_sym_10nm}(a) one can identify the corner states as the lower 
group of states, separated from the other states by an energy interval of $\Delta=1.2$ meV.  
However, the corner peaks spread into the facets, where they overlap. 
This effect is the strongest for the ground state, Fig.\ \ref{fig_sym_10nm}(b), and with
increasing the energy the maxima sharpen and the probability of finding an 
electron in the middle of a side decreases, Figs.\ \ref{fig_sym_10nm}(c) 
and\ \ref{fig_sym_10nm}(d), or even vanishes for the highest level in the corner group, 
Fig.\ \ref{fig_sym_10nm}(e). The second group is built of 12 states associated with 
probability distributions which form one maximum on each side. In the case of 
the lowest level, the localization pattern consists of only 6 maxima and electrons 
with this energy are depleted from corner areas, Fig.\ \ref{fig_sym_10nm}(f). For the 
higher states another 6 peaks appear in the vicinity of the vertices. Although they considerably 
increase with the energy, they never reach the height of the side maxima, 
Figs.\ \ref{fig_sym_10nm}(g)-\ \ref{fig_sym_10nm}(i). The number of maxima formed on 
each side increases by one between consecutive groups of 12 states, and thus 
for the following groups on the energy scale there are 2 and 3 maxima on every side,
respectively (not shown).

\begin{figure}[h]
\centering
 \includegraphics[scale=0.093]{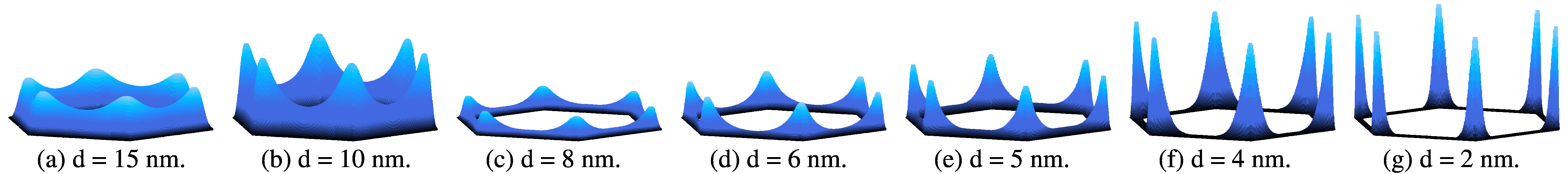} 
\caption{Probability distributions corresponding to the ground state of a single
         electron confined in symmetric rings restricted by the external radius 
         of 60 nm versus side thickness $d$.
         }
\label{fig_thickness}
\end{figure}

The shape of the probability distributions is governed by the aspect
ratio between the side 
thickness and the external radius ($d/R_{\mathrm{ext}}$). If it is large, i.e., for 
thick rings, the corner maxima penetrate deeply into the sides and strongly overlap
there, Figs.\ \ref{fig_thickness}(a) and\ \ref{fig_thickness}(b). Such samples 
do not differ much from circular ones, since in both cases the electrons may occupy
the whole circumference and easily rotate around the ring. 
If the corners are rounded, or destroyed, as in a more realistic situation, 
then the corner localization softens \cite{Sitek15}, 
and such samples resemble the circular rings even more. While the corner maxima sharpen
with decreasing the aspect ratio, they still overlap in the middle of the sides 
for a wide range of ring widths, Fig.\ \ref{fig_thickness}, 
except for very thin structures, 
Figs.\ \ref{fig_thickness}(f) and\ \ref{fig_thickness}(g).  

\begin{figure}[h]
\centering
 \includegraphics[scale=0.093]{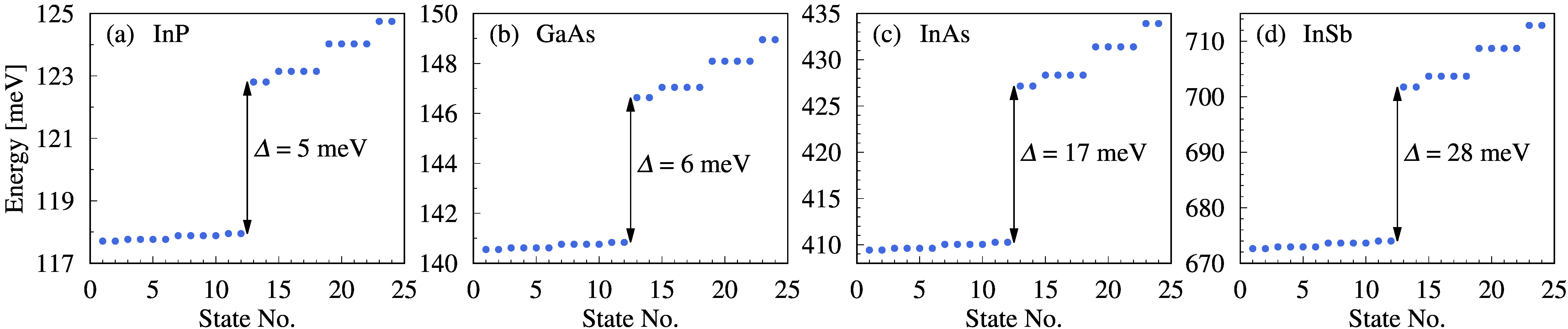} 
\caption{Energy levels of a single electron confined in symmetric quantum rings 
         of different materials. The side thickness and external radius are equal
         to 6 and 60 nm, respectively. $\Delta$ stands for the gap between corner 
         and side states.}
\label{fig_sym_6nm}
\end{figure}

The energy spacings between adjacent energy levels, and in particular the gap separating 
the corner from side states ($\Delta$), 
strongly depend on the aspect ratio ($d/R_{\mathrm{ext}}$) and the diameter 
($2R_{\mathrm{ext}}$). For the 10 nm thick InP ring with the external diameter of 120 nm $\Delta$
is equal to 1.2 meV, Fig.\ \ref{fig_sym_10nm}(a).
Decreasing the side thickness by 4 nm, while the diameter is kept constant, 
results in
an increase of over 4 times  
of this gap, Fig.\ \ref{fig_sym_6nm}(a). Even though $\Delta$ becomes thereby the dominant 
energy gap, which exceeds the dispersion of the corner states, it is still much smaller 
than for a triangular ring of the same thickness and external radius \cite{Sitek15}. 
The energy spectrum scales with the effective mass ($m_{\mathrm{eff}}$), 
i.e., the smaller the effective mass of the material is the larger the separation of the 
corner states and their energy dispersion are. In Fig.\ \ref{fig_sym_6nm} we compare the energy
spectra for four rings made of different semiconducting materials. A substantial fraction 
of experiments was performed on core-shell nanowires with the shell made of InP or GaAs.
Both of these materials are characterized by relatively large values of the effective mass, 
Table\ \ref{tab}, and thus even though for the 6 nm thick ring the gap between corner and side 
states is considerably larger than other energy intervals in the system, it is still below the 
resolution of most experiments, Figs.\ \ref{fig_sym_6nm}(a) and\ \ref{fig_sym_6nm}(b).
Still, $\Delta$ increases to 17 meV for InAs, Fig.\ \ref{fig_sym_6nm}(c), and for an InSb 
ring of the same shape it becomes 28 meV, Fig.\ \ref{fig_sym_6nm}(d).

\subsection{\label{Sym_many} Many-body states}

\begin{figure}[h]
\centering
 \includegraphics[scale=0.10]{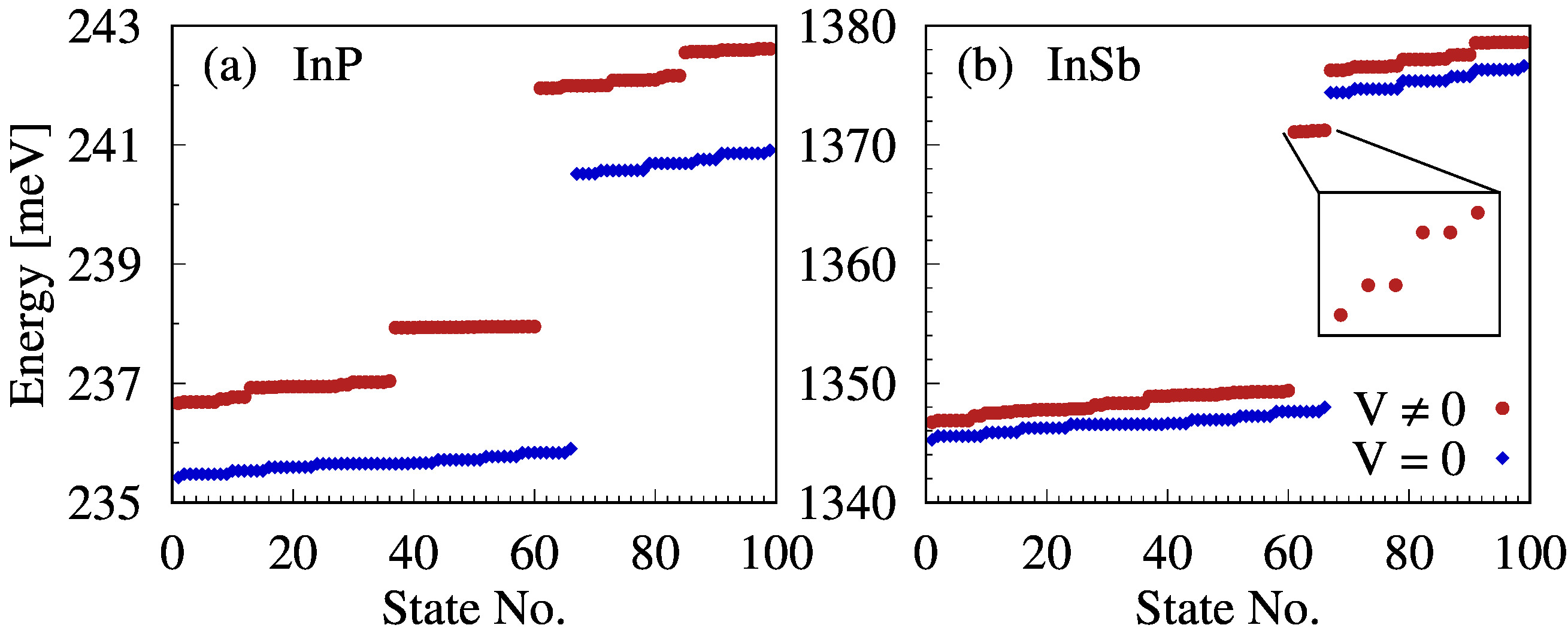}  
\caption{Energy levels of a pair of electrons confined in InP (a) and InSb (b) quantum rings
	     of 6 nm thickness.
         The blue diamonds correspond to noninteracting electrons, while the red circles represent Coulomb
         interacting particles. In the inset to Fig. b we show the degeneracy of the in-gap states.      }
\label{fig_many}
\end{figure}

The Coulomb repulsion between electrons adds to the energy of the quantum states. A question
is how this energy compares to the energy $\Delta$. If $\Delta$ is smaller than the characteristic Coulomb energy, many-body corner and side states will mix up. 
On the contrary, if $\Delta$ is the largest energy, then pure corner many body states 
may have energies within this gap \cite{Sitek17}.  In the example shown in Fig.\ \ref{fig_many}, 
66 eigenstates of two noninteracting electrons are separated 
from the higher states by approximately $\Delta$ (blue diamonds). The
Coulomb interaction shifts all states to higher energies and rearranges them according to 
the particle distribution within the ring. In particular, the lowest states correspond to 
particles localized around the opposite or alternating corners, respectively. The 24 states 
representing electrons occupying neighboring corners form a quasidegenerate level split  
from other corner states by a small gap, resulting from the decreased spatial separation of 
the particles. The number of separated corner states of the interacting system is reduced 
by 6 with respect to the noninteracting case.  The remaining states correspond to pairs of 
electrons in a spin singlet configuration accumulated in the same corner area for which the 
contribution due to the 
electrostatic repulsion exceeds $\Delta$, and thus these states mix with the states above 
the gap [red circles in Fig.\ \ref{fig_many}(a)]. The splitting between 
single-particle corner and side states increases for materials with smaller effective mass. 
Although, for the set of studied materials the relative permittivity increases (Table\ \ref{tab}),
and thus the contributions due to Coulomb interaction decrease.  Consequently, for 
the InSb ring the gap $\Delta$ is
larger than the Coulomb-induced shift of the pairs of close-by electrons. As a result, 
the corresponding six states stay 
below the states associated with mixed corner-side probability distributions, i.e., in the 
gap, Fig.\ \ref{fig_many}(b). Such in-gap states were previously obtained for triangular 
rings where the gap $\Delta$ is much larger than in the hexagonal case \cite{Sitek17}.
In principle, such states should also appear for ultra thin InP and GaAs hexagonal 
shells, but such rings are beyond our computational limitations. 

\section{\label{Def} Deformed rings}

\subsection{\label{Def_asym} Side thickness and corner deformations}

In spite of high-precision manufacturing technologies it is still
impossible to obtain perfectly symmetric wires, and it is even more
difficult to cover the wires with shells of constant thickness.  Such
wires are grown in sets of close-by vertical cores which are later covered
with layers of a different material. Due to the screening of neighboring
wires, the shell thickness varies along the cross section circumference
\cite{Battiato19,Sonner19}.  
In general, the corner localization is very sensitive to the ring 
symmetry and to the size of the corner area \cite{Sitek16}. 
Ballester \textit{et al.} \cite{Ballester12} 
analyzed a hexagonal ring with one thicker side and
showed that the lowest single-particle states are localized on that 
side. Starting with a symmetric hexagon and increasing the thickness of one side,
the probability distribution corresponding to the ground state forms initially 
two maxima at the ends of the thickest side, which further merge to form one maximum 
along the whole facet. The number of states localized on the widest side depends on the
ratio between the thickness of this side and the width of the others. The
following group on the energy scale consists of four corner states,
associated with probability distributions forming four peaks localized
around the corner areas of the same size \cite{Ballester12}.

\begin{figure}[h]
\centering
 \includegraphics[scale=0.10]{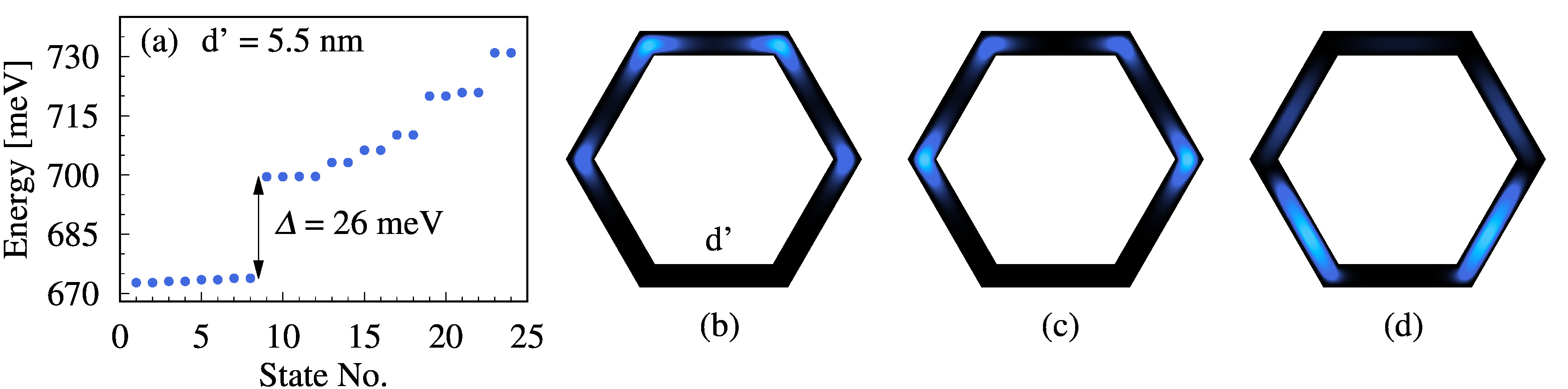}  
\caption{Single-particle energies for a ring with one thinner side [side $d'$ in Fig. (b)] (a),
         and the examples of the corresponding probability distributions [(b)-(d)].} 
\label{fig_asym}
\end{figure}

We consider the opposite case, i.e., the situation when one of the sides
is thinner than the other ones. Here the electrons are depleted from the
thinner side and the structure acts as a system with five facets. The
low-energy states are distributed between the four larger corner areas,
while the electrons excited to the higher levels are delocalized over the
five sides.  Obviously, the probability distributions
corresponding to both types of states do not reproduce the symmetry of the
sample, but only the mirror symmetry with respect to the thinnest
side, Figs.\ \ref{fig_asym}(b)-\ref{fig_asym}(d).  As for square rings
\cite{Sitek15}, there are eight corner-localized states, spread within a
narrow energy range and separated from the ten higher side-localized
states, i.e., the corner states are still protected by a considerable
energy gap, but slightly smaller than in the case of a square ring of
the same diameter and width, Fig.\ \ref{fig_asym}(a).

\begin{figure}[h]
\centering
 \includegraphics[scale=0.10]{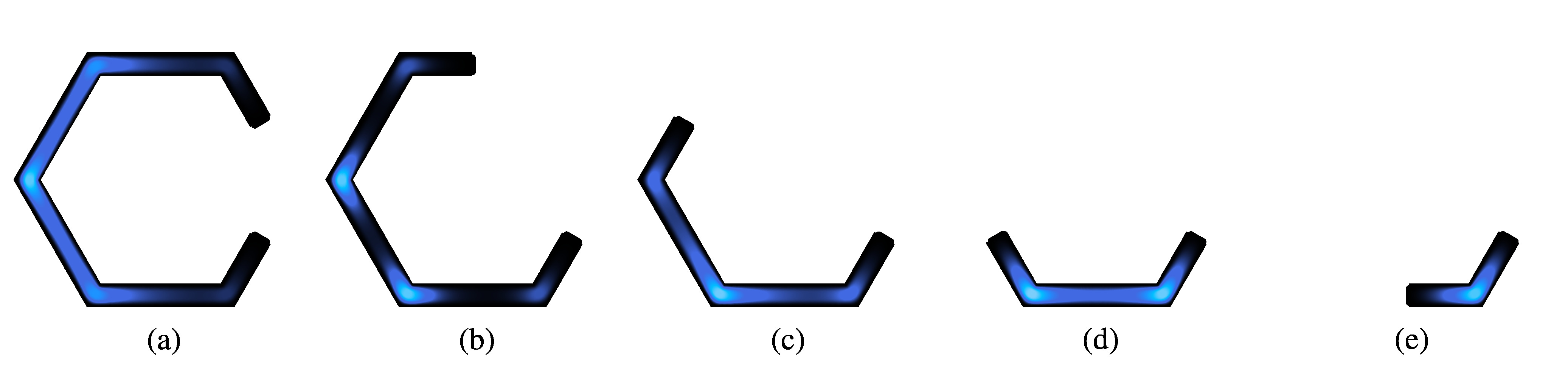}  
\caption{Ground state probability distributions for quantum rings with reduced number of corners.} 
\label{fig_loc}
\end{figure}

\begin{figure}[h]
\centering
 \includegraphics[scale=0.10]{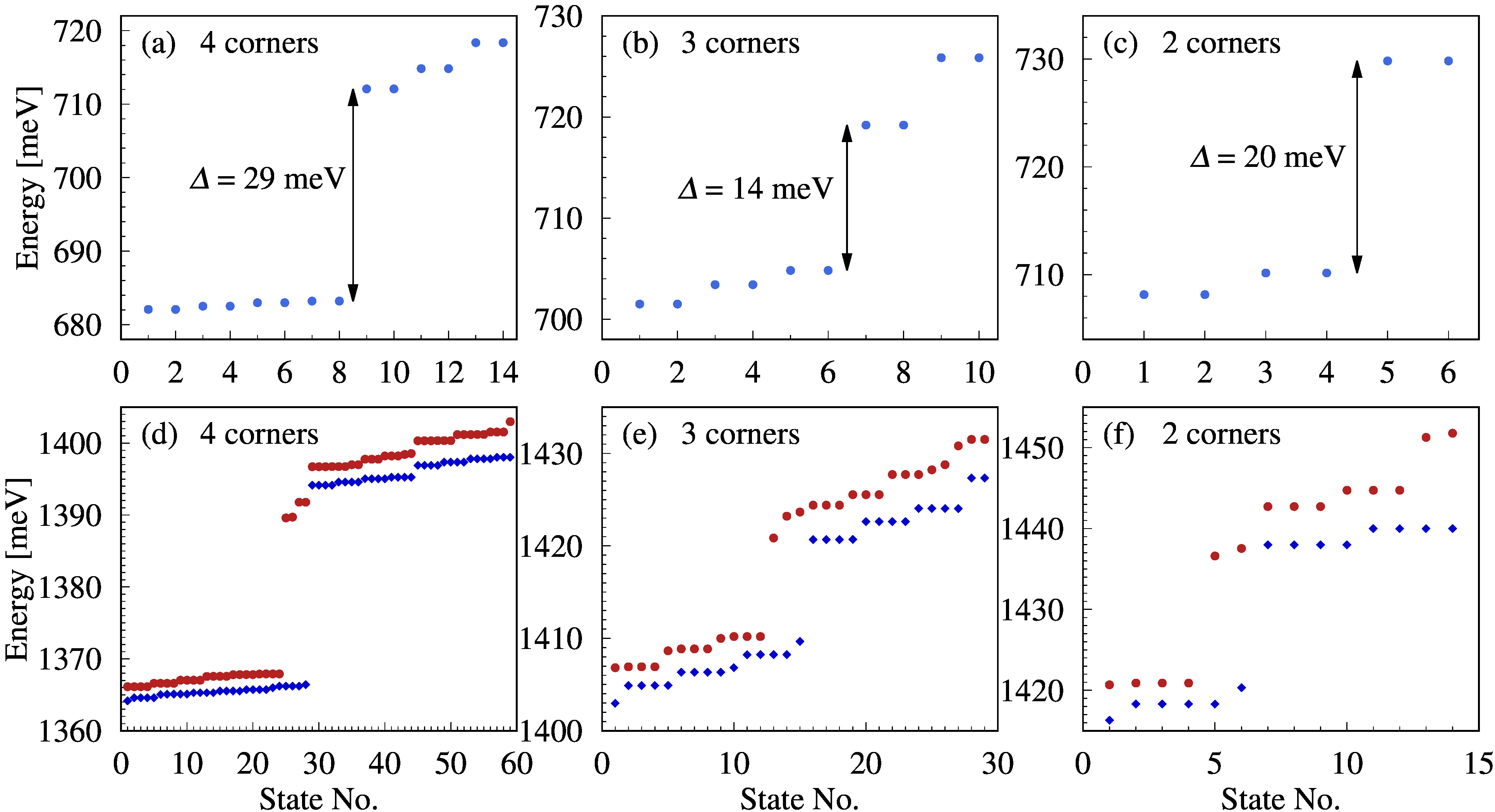}  
\caption{Single-particle [(a)-(c)] and many-body [(d)-(f)] energies of two 
         particles confined in quantum rings with reduced number of corners.} 
\label{fig_cut}
\end{figure}

During the fabrication process, the coverage of the core with the shell material may be 
nonuniform, such that some regions of the shell may be very thin or even absent \cite{Yuan15,Guniat19}.
We illustrate such examples in Fig.\ \ref{fig_loc}.
In such cases, the lowest
energy levels are also associated with corner-localized probability
distributions. 
The corner localization is not
surprising since such structures resemble multibent wires and it had
already been shown that in the bent parts of such structures effective
quantum wells are formed \cite{Wu}. Interestingly, the separation of the
corner states from the side ones initially, i.e., when one of the corners
is excluded, drops down and later increases and fluctuates with the subsequent reduction
of the number of corners, Figs.\ \ref{fig_cut}(a)-\ref{fig_cut}(c).
More importantly, if the shell is made of InSb material, then the
in-gap states are obtained down to the two-corner system, Figs.\
\ref{fig_cut}(d)-\ref{fig_cut}(f).

\subsection{\label{Def_sym} Angular deformations}

\begin{figure}[h]
\centering
\includegraphics[scale=0.10]{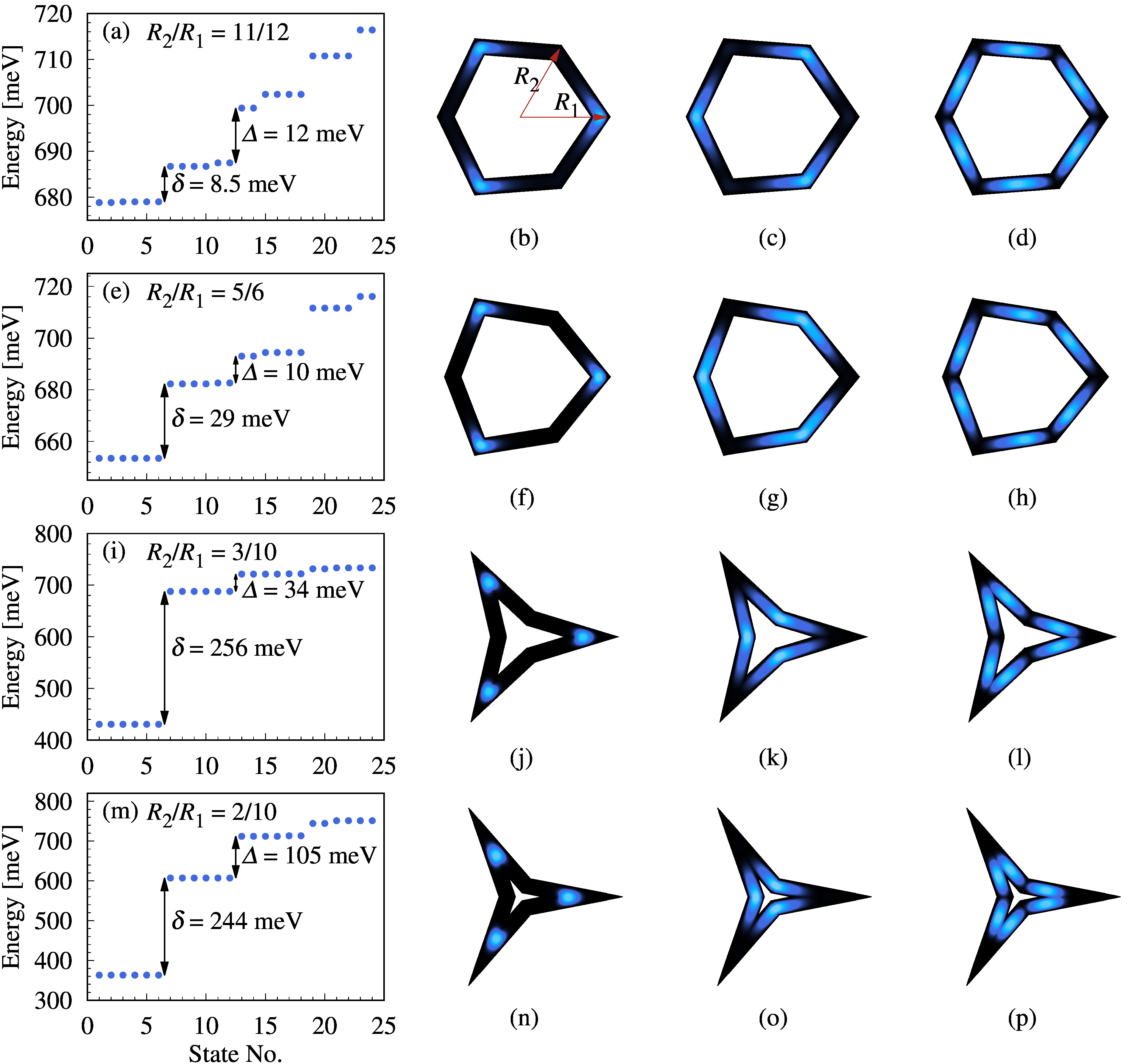}  
\caption{Single-particle energies for InSb rings with different ratios of the neighboring
         external radii $R_2/R_1$ (first column), and the probability distributions 
         corresponding to the ground state (second column), the second corner level 
         (third column), and the lowest side level (fourth column).}
\label{fig_stars}
\end{figure}
\begin{figure}[h]
\centering
 \includegraphics[scale=0.12]{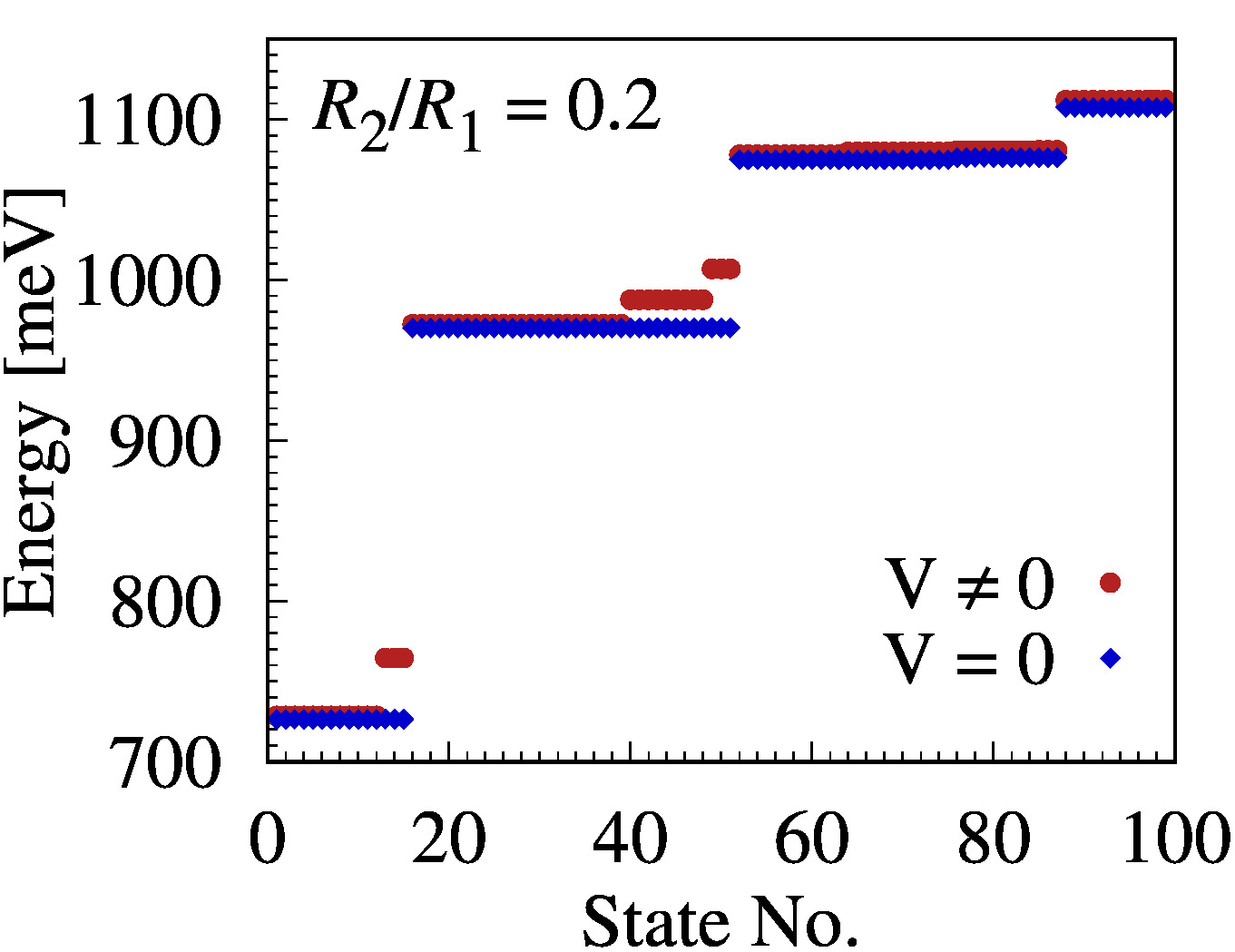}  
\caption{Energy levels for a pair of electrons confined in 
         an InSb ring for which $R_2/R_1=0.2$.}
\label{fig_star_E_many}
\end{figure}

Next, we consider hexagonal rings with angular deformations. 
This class includes star shaped polygons which have been recently obtained \cite{Ameruddin15}. 
We assume that three
alternating external radii ($R_1$) do not change, while the other three are
reduced ($R_2$), Fig.\ \ref{fig_stars}. 
As a result, the system is now threefold rotationally 
symmetric, i.e., it is built of three identical elements.
The two different angles split the corner domain into two 
groups of corner states \cite{Sitek15ICTON}, 
separated by a gap ($\delta$), which is comparable to the separation of the 
corner states from side states ($\Delta$) for small differences between the two radii, Fig.\ \ref{fig_stars}(a). 
The probability distributions corresponding to the lowest states are localized around the sharpest
corners, Fig.\ \ref{fig_stars}(b), while those associated with
the second group of corner states form three maxima residing in the larger corner areas 
Fig.\ \ref{fig_stars}(c). 
These, relatively small, differences of the corner areas do not affect the side localization, 
which remains symmetric, Fig.\ \ref{fig_stars}(d). Decreasing the ratio $R_2/R_1$ results in the sharpening of 
three corner areas and softening of the alternating ones. This considerably increases the separation  
between the corner states ($\delta$) such that it becomes the dominant gap of the spectrum, while 
the separation of the corner from side states slightly decreases, Fig.\ \ref{fig_stars}(e). The lowest 
states are strongly localized
in the vicinity of the sharpest corners, Fig.\ \ref{fig_stars}(f), while the probability distributions 
associated with the second group of corner states form elongated maxima centered around the 
bent parts, Fig.\ \ref{fig_stars}(g). 
In this case the side-localized maxima are shifted towards the softer corners, Fig.\ \ref{fig_stars}(h). 
The decrease of the ratio $R_2/R_1$ results in obtaining triangular quantum rings which were 
analyzed elsewhere \cite{Sitek15,Sitek16,Sitek17,Sitek18}, and star-like structures.  
For the latter ones the gap separating the corner from side states ($\Delta$) slightly increases, 
but the lower gap ($\delta$) exceeds it by nearly one order of magnitude, Fig.\ \ref{fig_stars}(i). 
The ratio of external radii ($R_2/R_1$) which is lower than that of a triangle (0.5), 
induces the formation of six  quasidegenerate corner-states. These states are well-separated from the 
higher, formally of corner-type, states but spatially elongated similarly to the side states 
of triangular rings. Further decrease of the ratio $R_2/R_1$ leads to the reduction of the 
wider corner areas, and thus sharpening of the corresponding localization peaks, Fig.\ \ref{fig_stars}(o). 
This results in the increase of the separation of this states from the side-localized states, Fig.\ \ref{fig_stars}(m), 
but does not affect much the localization of the ground state and the side-states, 
Figs.\ \ref{fig_stars}(n) and\ \ref{fig_stars}(p), respectively. Similarly to triangular rings \cite{Sitek17}
the star-like shells allow to obtain 3 in-gap states in the largest gap and on top of that, 
they enable the formation of well-separated 
levels in the second gap of the many-body spectrum of a pair of Coulomb coupled carriers, 
Fig.\ \ref{fig_star_E_many}.

\section{\label{Con} Conclusions}

We studied the energy levels and localization of single and two Coulomb interacting electrons 
confined in symmetric and nonsymmetric hexagonal quantum rings made of commonly used semiconductor 
materials: InP, GaAs, InAs, and InSb.   The results that we showed are relevant not only for quantum rings, 
but also for the transverse modes in long nanowires of core-shell type.  

The wave functions of the low energy 
states are distributed either in the corner areas or on the sides of the hexagonal shell.  The separation of 
the corner and side states, both in terms of localization and in energy, depends on the geometry parameters 
and on the material, via the effective mass.   In particular, with InSb for which the effective mass is the 
smallest, the single-particle corner states may be separated from the side-localized states by a gap of 28 meV, 
i.e., comparable to the room-temperature energy, for a ring of 60 nm radius and 6 nm thickness.   
In this case, 
the many-body energy spectrum resembles the one of triangular rings, i.e., the lowest states 
are spread within a small energy range and are followed by in-gap states.

The separation of 
the in-gap states from the lower corner states, and also from the higher corner-side states, 
may considerably exceed the energy dispersion 
of the lowest group of corner sates. External electric and magnetic fields may easily change the arrangement of the latter states, and thus the spin configuration of the ground state, without 
altering much the in-gap states. Since all in-gap states are of the spin-singlet type, 
in the absence of spin-orbit interaction they may be optically excited only from a state of the 
same spin configuration. 
This allows to block the excitation of these states in the presence 
of an external field which changes the ground state, i.e., 
as for the triangular rings, such energy structure allows for the 
contactless control of absorption \cite{Sitek17}.  This aspect is 
interesting especially from the technological point of view, because it is much easier 
to obtain imperfect hexagonal structures than triangular.  

Finally, we show that in the star-shaped hexagons the lowest energy states are strongly localized in the sharp corners, an effect similar to the quantum localization in deep quantum wells, and that these states can be separated from the other ones by a gap which exceeds the room-temperature energy by approximately one order 
of magnitude.

\begin{acknowledgments}
This work was supported by the Icelandic Research Fund.
\end{acknowledgments}

\bibliography{core_shell}

\begin{thebibliography}{43}%
\makeatletter
\providecommand \@ifxundefined [1]{%
 \@ifx{#1\undefined}
}%
\providecommand \@ifnum [1]{%
 \ifnum #1\expandafter \@firstoftwo
 \else \expandafter \@secondoftwo
 \fi
}%
\providecommand \@ifx [1]{%
 \ifx #1\expandafter \@firstoftwo
 \else \expandafter \@secondoftwo
 \fi
}%
\providecommand \natexlab [1]{#1}%
\providecommand \enquote  [1]{``#1''}%
\providecommand \bibnamefont  [1]{#1}%
\providecommand \bibfnamefont [1]{#1}%
\providecommand \citenamefont [1]{#1}%
\providecommand \href@noop [0]{\@secondoftwo}%
\providecommand \href [0]{\begingroup \@sanitize@url \@href}%
\providecommand \@href[1]{\@@startlink{#1}\@@href}%
\providecommand \@@href[1]{\endgroup#1\@@endlink}%
\providecommand \@sanitize@url [0]{\catcode `\\12\catcode `\$12\catcode
  `\&12\catcode `\#12\catcode `\^12\catcode `\_12\catcode `\%12\relax}%
\providecommand \@@startlink[1]{}%
\providecommand \@@endlink[0]{}%
\providecommand \url  [0]{\begingroup\@sanitize@url \@url }%
\providecommand \@url [1]{\endgroup\@href {#1}{\urlprefix }}%
\providecommand \urlprefix  [0]{URL }%
\providecommand \Eprint [0]{\href }%
\providecommand \doibase [0]{http://dx.doi.org/}%
\providecommand \selectlanguage [0]{\@gobble}%
\providecommand \bibinfo  [0]{\@secondoftwo}%
\providecommand \bibfield  [0]{\@secondoftwo}%
\providecommand \translation [1]{[#1]}%
\providecommand \BibitemOpen [0]{}%
\providecommand \bibitemStop [0]{}%
\providecommand \bibitemNoStop [0]{.\EOS\space}%
\providecommand \EOS [0]{\spacefactor3000\relax}%
\providecommand \BibitemShut  [1]{\csname bibitem#1\endcsname}%
\let\auto@bib@innerbib\@empty
\bibitem [{\citenamefont {Bl{\"o}mers}\ \emph {et~al.}(2013)\citenamefont
  {Bl{\"o}mers}, \citenamefont {Rieger}, \citenamefont {Zellekens},
  \citenamefont {Haas}, \citenamefont {Lepsa}, \citenamefont {Hardtdegen},
  \citenamefont {G{\"u}l}, \citenamefont {Demarina}, \citenamefont
  {Gr{\"u}tzmacher}, \citenamefont {L{\"u}th},\ and\ \citenamefont
  {Sch{\"a}pers}}]{Blomers13}%
  \BibitemOpen
  \bibfield  {author} {\bibinfo {author} {\bibfnamefont {C.}~\bibnamefont
  {Bl{\"o}mers}}, \bibinfo {author} {\bibfnamefont {T.}~\bibnamefont {Rieger}},
  \bibinfo {author} {\bibfnamefont {P.}~\bibnamefont {Zellekens}}, \bibinfo
  {author} {\bibfnamefont {F.}~\bibnamefont {Haas}}, \bibinfo {author}
  {\bibfnamefont {M.~I.}\ \bibnamefont {Lepsa}}, \bibinfo {author}
  {\bibfnamefont {H.}~\bibnamefont {Hardtdegen}}, \bibinfo {author}
  {\bibfnamefont {{\"O}.}~\bibnamefont {G{\"u}l}}, \bibinfo {author}
  {\bibfnamefont {N.}~\bibnamefont {Demarina}}, \bibinfo {author}
  {\bibfnamefont {D.}~\bibnamefont {Gr{\"u}tzmacher}}, \bibinfo {author}
  {\bibfnamefont {H.}~\bibnamefont {L{\"u}th}}, \ and\ \bibinfo {author}
  {\bibfnamefont {T.}~\bibnamefont {Sch{\"a}pers}},\ }\href@noop {} {\bibfield
  {journal} {\bibinfo  {journal} {Nanotechnology}\ }\textbf {\bibinfo {volume}
  {24}},\ \bibinfo {pages} {035203} (\bibinfo {year} {2013})}\BibitemShut
  {NoStop}%
\bibitem [{\citenamefont {Rieger}\ \emph {et~al.}(2012)\citenamefont {Rieger},
  \citenamefont {Luysberg}, \citenamefont {Sch{\"a}pers}, \citenamefont
  {Gr{\"u}tzmacher},\ and\ \citenamefont {Lepsa}}]{Rieger12}%
  \BibitemOpen
  \bibfield  {author} {\bibinfo {author} {\bibfnamefont {T.}~\bibnamefont
  {Rieger}}, \bibinfo {author} {\bibfnamefont {M.}~\bibnamefont {Luysberg}},
  \bibinfo {author} {\bibfnamefont {T.}~\bibnamefont {Sch{\"a}pers}}, \bibinfo
  {author} {\bibfnamefont {D.}~\bibnamefont {Gr{\"u}tzmacher}}, \ and\ \bibinfo
  {author} {\bibfnamefont {M.~I.}\ \bibnamefont {Lepsa}},\ }\href@noop {}
  {\bibfield  {journal} {\bibinfo  {journal} {Nano Letters}\ }\textbf {\bibinfo
  {volume} {12}},\ \bibinfo {pages} {5559} (\bibinfo {year}
  {2012})}\BibitemShut {NoStop}%
\bibitem [{\citenamefont {Haas}\ \emph {et~al.}(2013)\citenamefont {Haas},
  \citenamefont {Sladek}, \citenamefont {Winden}, \citenamefont {von~der Ahe},
  \citenamefont {Weirich}, \citenamefont {Rieger}, \citenamefont {L{\"u}th},
  \citenamefont {Gr{\"u}tzmacher}, \citenamefont {Sch{\"a}pers},\ and\
  \citenamefont {Hardtdegen}}]{Haas13}%
  \BibitemOpen
  \bibfield  {author} {\bibinfo {author} {\bibfnamefont {F.}~\bibnamefont
  {Haas}}, \bibinfo {author} {\bibfnamefont {K.}~\bibnamefont {Sladek}},
  \bibinfo {author} {\bibfnamefont {A.}~\bibnamefont {Winden}}, \bibinfo
  {author} {\bibfnamefont {M.}~\bibnamefont {von~der Ahe}}, \bibinfo {author}
  {\bibfnamefont {T.~E.}\ \bibnamefont {Weirich}}, \bibinfo {author}
  {\bibfnamefont {T.}~\bibnamefont {Rieger}}, \bibinfo {author} {\bibfnamefont
  {H.}~\bibnamefont {L{\"u}th}}, \bibinfo {author} {\bibfnamefont
  {D.}~\bibnamefont {Gr{\"u}tzmacher}}, \bibinfo {author} {\bibfnamefont
  {T.}~\bibnamefont {Sch{\"a}pers}}, \ and\ \bibinfo {author} {\bibfnamefont
  {H.}~\bibnamefont {Hardtdegen}},\ }\href@noop {} {\bibfield  {journal}
  {\bibinfo  {journal} {Nanotechnology}\ }\textbf {\bibinfo {volume} {24}},\
  \bibinfo {pages} {085603} (\bibinfo {year} {2013})}\BibitemShut {NoStop}%
\bibitem [{\citenamefont {Funk}\ \emph {et~al.}(2013)\citenamefont {Funk},
  \citenamefont {Royo}, \citenamefont {Zardo}, \citenamefont {Rudolph},
  \citenamefont {Morkötter}, \citenamefont {Mayer}, \citenamefont {Becker},
  \citenamefont {Bechtold}, \citenamefont {Matich}, \citenamefont {Döblinger},
  \citenamefont {Bichler}, \citenamefont {Koblmüller}, \citenamefont {Finley},
  \citenamefont {Bertoni}, \citenamefont {Goldoni},\ and\ \citenamefont
  {Abstreiter}}]{Funk13}%
  \BibitemOpen
  \bibfield  {author} {\bibinfo {author} {\bibfnamefont {S.}~\bibnamefont
  {Funk}}, \bibinfo {author} {\bibfnamefont {M.}~\bibnamefont {Royo}}, \bibinfo
  {author} {\bibfnamefont {I.}~\bibnamefont {Zardo}}, \bibinfo {author}
  {\bibfnamefont {D.}~\bibnamefont {Rudolph}}, \bibinfo {author} {\bibfnamefont
  {S.}~\bibnamefont {Morkötter}}, \bibinfo {author} {\bibfnamefont
  {B.}~\bibnamefont {Mayer}}, \bibinfo {author} {\bibfnamefont
  {J.}~\bibnamefont {Becker}}, \bibinfo {author} {\bibfnamefont
  {A.}~\bibnamefont {Bechtold}}, \bibinfo {author} {\bibfnamefont
  {S.}~\bibnamefont {Matich}}, \bibinfo {author} {\bibfnamefont
  {M.}~\bibnamefont {Döblinger}}, \bibinfo {author} {\bibfnamefont
  {M.}~\bibnamefont {Bichler}}, \bibinfo {author} {\bibfnamefont
  {G.}~\bibnamefont {Koblmüller}}, \bibinfo {author} {\bibfnamefont {J.~J.}\
  \bibnamefont {Finley}}, \bibinfo {author} {\bibfnamefont {A.}~\bibnamefont
  {Bertoni}}, \bibinfo {author} {\bibfnamefont {G.}~\bibnamefont {Goldoni}}, \
  and\ \bibinfo {author} {\bibfnamefont {G.}~\bibnamefont {Abstreiter}},\
  }\href@noop {} {\bibfield  {journal} {\bibinfo  {journal} {Nano Letters}\
  }\textbf {\bibinfo {volume} {13}},\ \bibinfo {pages} {6189} (\bibinfo {year}
  {2013})}\BibitemShut {NoStop}%
\bibitem [{\citenamefont {Erhard}\ \emph {et~al.}(2015)\citenamefont {Erhard},
  \citenamefont {Zenger}, \citenamefont {Morkötter}, \citenamefont {Rudolph},
  \citenamefont {Weiss}, \citenamefont {Krenner}, \citenamefont {Karl},
  \citenamefont {Abstreiter}, \citenamefont {Finley}, \citenamefont
  {Koblm{\"u}ller},\ and\ \citenamefont {Holleitner}}]{Erhard15}%
  \BibitemOpen
  \bibfield  {author} {\bibinfo {author} {\bibfnamefont {N.}~\bibnamefont
  {Erhard}}, \bibinfo {author} {\bibfnamefont {S.}~\bibnamefont {Zenger}},
  \bibinfo {author} {\bibfnamefont {S.}~\bibnamefont {Morkötter}}, \bibinfo
  {author} {\bibfnamefont {D.}~\bibnamefont {Rudolph}}, \bibinfo {author}
  {\bibfnamefont {M.}~\bibnamefont {Weiss}}, \bibinfo {author} {\bibfnamefont
  {H.~J.}\ \bibnamefont {Krenner}}, \bibinfo {author} {\bibfnamefont
  {H.}~\bibnamefont {Karl}}, \bibinfo {author} {\bibfnamefont {G.}~\bibnamefont
  {Abstreiter}}, \bibinfo {author} {\bibfnamefont {J.~J.}\ \bibnamefont
  {Finley}}, \bibinfo {author} {\bibfnamefont {G.}~\bibnamefont
  {Koblm{\"u}ller}}, \ and\ \bibinfo {author} {\bibfnamefont {A.~W.}\
  \bibnamefont {Holleitner}},\ }\href@noop {} {\bibfield  {journal} {\bibinfo
  {journal} {Nano Letters}\ }\textbf {\bibinfo {volume} {15}},\ \bibinfo
  {pages} {6869} (\bibinfo {year} {2015})}\BibitemShut {NoStop}%
\bibitem [{\citenamefont {Wei{\ss}}\ \emph {et~al.}(2014)\citenamefont
  {Wei{\ss}}, \citenamefont {Kinzel}, \citenamefont {Sch{\"u}lein},
  \citenamefont {Heigl}, \citenamefont {Rudolph}, \citenamefont
  {Mork{\"o}tter}, \citenamefont {D{\"o}blinger}, \citenamefont {Bichler},
  \citenamefont {Abstreiter}, \citenamefont {Finley}, \citenamefont
  {Koblm{\"u}ller}, \citenamefont {Wixforth},\ and\ \citenamefont
  {Krenner}}]{Weiss14b}%
  \BibitemOpen
  \bibfield  {author} {\bibinfo {author} {\bibfnamefont {M.}~\bibnamefont
  {Wei{\ss}}}, \bibinfo {author} {\bibfnamefont {J.~B.}\ \bibnamefont
  {Kinzel}}, \bibinfo {author} {\bibfnamefont {F.~J.~R.}\ \bibnamefont
  {Sch{\"u}lein}}, \bibinfo {author} {\bibfnamefont {M.}~\bibnamefont {Heigl}},
  \bibinfo {author} {\bibfnamefont {D.}~\bibnamefont {Rudolph}}, \bibinfo
  {author} {\bibfnamefont {S.}~\bibnamefont {Mork{\"o}tter}}, \bibinfo {author}
  {\bibfnamefont {M.}~\bibnamefont {D{\"o}blinger}}, \bibinfo {author}
  {\bibfnamefont {M.}~\bibnamefont {Bichler}}, \bibinfo {author} {\bibfnamefont
  {G.}~\bibnamefont {Abstreiter}}, \bibinfo {author} {\bibfnamefont {J.~J.}\
  \bibnamefont {Finley}}, \bibinfo {author} {\bibfnamefont {G.}~\bibnamefont
  {Koblm{\"u}ller}}, \bibinfo {author} {\bibfnamefont {A.}~\bibnamefont
  {Wixforth}}, \ and\ \bibinfo {author} {\bibfnamefont {H.~J.}\ \bibnamefont
  {Krenner}},\ }\href@noop {} {\bibfield  {journal} {\bibinfo  {journal} {Nano
  Letters}\ }\textbf {\bibinfo {volume} {14}},\ \bibinfo {pages} {2256}
  (\bibinfo {year} {2014})}\BibitemShut {NoStop}%
\bibitem [{\citenamefont {Jadczak}\ \emph {et~al.}(2014)\citenamefont
  {Jadczak}, \citenamefont {Plochocka}, \citenamefont {Mitioglu}, \citenamefont
  {Breslavetz}, \citenamefont {Royo}, \citenamefont {Bertoni}, \citenamefont
  {Goldoni}, \citenamefont {Smolenski}, \citenamefont {Kossacki}, \citenamefont
  {Kretinin}, \citenamefont {Shtrikman},\ and\ \citenamefont
  {Maude}}]{Jadczak14}%
  \BibitemOpen
  \bibfield  {author} {\bibinfo {author} {\bibfnamefont {J.}~\bibnamefont
  {Jadczak}}, \bibinfo {author} {\bibfnamefont {P.}~\bibnamefont {Plochocka}},
  \bibinfo {author} {\bibfnamefont {A.}~\bibnamefont {Mitioglu}}, \bibinfo
  {author} {\bibfnamefont {I.}~\bibnamefont {Breslavetz}}, \bibinfo {author}
  {\bibfnamefont {M.}~\bibnamefont {Royo}}, \bibinfo {author} {\bibfnamefont
  {A.}~\bibnamefont {Bertoni}}, \bibinfo {author} {\bibfnamefont
  {G.}~\bibnamefont {Goldoni}}, \bibinfo {author} {\bibfnamefont
  {T.}~\bibnamefont {Smolenski}}, \bibinfo {author} {\bibfnamefont
  {P.}~\bibnamefont {Kossacki}}, \bibinfo {author} {\bibfnamefont
  {A.}~\bibnamefont {Kretinin}}, \bibinfo {author} {\bibfnamefont
  {H.}~\bibnamefont {Shtrikman}}, \ and\ \bibinfo {author} {\bibfnamefont
  {D.~K.}\ \bibnamefont {Maude}},\ }\href@noop {} {\bibfield  {journal}
  {\bibinfo  {journal} {Nano Letters}\ }\textbf {\bibinfo {volume} {14}},\
  \bibinfo {pages} {2807} (\bibinfo {year} {2014})}\BibitemShut {NoStop}%
\bibitem [{\citenamefont {Qian}\ \emph {et~al.}(2004)\citenamefont {Qian},
  \citenamefont {Li}, \citenamefont {Grade{\v{c}}ak}, \citenamefont {Wang},
  \citenamefont {Barrelet},\ and\ \citenamefont {Lieber}}]{Qian04}%
  \BibitemOpen
  \bibfield  {author} {\bibinfo {author} {\bibfnamefont {F.}~\bibnamefont
  {Qian}}, \bibinfo {author} {\bibfnamefont {Y.}~\bibnamefont {Li}}, \bibinfo
  {author} {\bibfnamefont {S.}~\bibnamefont {Grade{\v{c}}ak}}, \bibinfo
  {author} {\bibfnamefont {D.}~\bibnamefont {Wang}}, \bibinfo {author}
  {\bibfnamefont {C.~J.}\ \bibnamefont {Barrelet}}, \ and\ \bibinfo {author}
  {\bibfnamefont {C.~M.}\ \bibnamefont {Lieber}},\ }\href@noop {} {\bibfield
  {journal} {\bibinfo  {journal} {Nano Letters}\ }\textbf {\bibinfo {volume}
  {4}},\ \bibinfo {pages} {1975} (\bibinfo {year} {2004})}\BibitemShut
  {NoStop}%
\bibitem [{\citenamefont {Qian}\ \emph {et~al.}(2005)\citenamefont {Qian},
  \citenamefont {Grade{\v{c}}ak}, \citenamefont {Li}, \citenamefont {Wen},\
  and\ \citenamefont {Lieber}}]{Qian05}%
  \BibitemOpen
  \bibfield  {author} {\bibinfo {author} {\bibfnamefont {F.}~\bibnamefont
  {Qian}}, \bibinfo {author} {\bibfnamefont {S.}~\bibnamefont
  {Grade{\v{c}}ak}}, \bibinfo {author} {\bibfnamefont {Y.}~\bibnamefont {Li}},
  \bibinfo {author} {\bibfnamefont {C.-Y.}\ \bibnamefont {Wen}}, \ and\
  \bibinfo {author} {\bibfnamefont {C.~M.}\ \bibnamefont {Lieber}},\
  }\href@noop {} {\bibfield  {journal} {\bibinfo  {journal} {Nano Letters}\
  }\textbf {\bibinfo {volume} {5}},\ \bibinfo {pages} {2287} (\bibinfo {year}
  {2005})}\BibitemShut {NoStop}%
\bibitem [{\citenamefont {Baird}\ \emph {et~al.}(2009)\citenamefont {Baird},
  \citenamefont {Ang}, \citenamefont {Low}, \citenamefont {Haegel},
  \citenamefont {Talin}, \citenamefont {Li},\ and\ \citenamefont
  {Wang}}]{Baird09}%
  \BibitemOpen
  \bibfield  {author} {\bibinfo {author} {\bibfnamefont {L.}~\bibnamefont
  {Baird}}, \bibinfo {author} {\bibfnamefont {G.}~\bibnamefont {Ang}}, \bibinfo
  {author} {\bibfnamefont {C.}~\bibnamefont {Low}}, \bibinfo {author}
  {\bibfnamefont {N.}~\bibnamefont {Haegel}}, \bibinfo {author} {\bibfnamefont
  {A.}~\bibnamefont {Talin}}, \bibinfo {author} {\bibfnamefont
  {Q.}~\bibnamefont {Li}}, \ and\ \bibinfo {author} {\bibfnamefont
  {G.}~\bibnamefont {Wang}},\ }\href@noop {} {\bibfield  {journal} {\bibinfo
  {journal} {Physica B: Condensed Matter}\ }\textbf {\bibinfo {volume} {404}},\
  \bibinfo {pages} {4933 } (\bibinfo {year} {2009})}\BibitemShut {NoStop}%
\bibitem [{\citenamefont {Heurlin}\ \emph {et~al.}(2015)\citenamefont
  {Heurlin}, \citenamefont {Stankevi{\v{c}}}, \citenamefont
  {Mickevi{\v{c}}ius}, \citenamefont {Yngman}, \citenamefont {Lindgren},
  \citenamefont {Mikkelsen}, \citenamefont {Feidenhans’l}, \citenamefont
  {Borgst{\"o}m},\ and\ \citenamefont {Samuelson}}]{Heurlin15}%
  \BibitemOpen
  \bibfield  {author} {\bibinfo {author} {\bibfnamefont {M.}~\bibnamefont
  {Heurlin}}, \bibinfo {author} {\bibfnamefont {T.}~\bibnamefont
  {Stankevi{\v{c}}}}, \bibinfo {author} {\bibfnamefont {S.}~\bibnamefont
  {Mickevi{\v{c}}ius}}, \bibinfo {author} {\bibfnamefont {S.}~\bibnamefont
  {Yngman}}, \bibinfo {author} {\bibfnamefont {D.}~\bibnamefont {Lindgren}},
  \bibinfo {author} {\bibfnamefont {A.}~\bibnamefont {Mikkelsen}}, \bibinfo
  {author} {\bibfnamefont {R.}~\bibnamefont {Feidenhans’l}}, \bibinfo
  {author} {\bibfnamefont {M.~T.}\ \bibnamefont {Borgst{\"o}m}}, \ and\
  \bibinfo {author} {\bibfnamefont {L.}~\bibnamefont {Samuelson}},\ }\href@noop
  {} {\bibfield  {journal} {\bibinfo  {journal} {Nano Letters}\ }\textbf
  {\bibinfo {volume} {15}},\ \bibinfo {pages} {2462} (\bibinfo {year}
  {2015})}\BibitemShut {NoStop}%
\bibitem [{\citenamefont {Dong}\ \emph {et~al.}(2009)\citenamefont {Dong},
  \citenamefont {Tian}, \citenamefont {Kempa},\ and\ \citenamefont
  {Lieber}}]{Dong09}%
  \BibitemOpen
  \bibfield  {author} {\bibinfo {author} {\bibfnamefont {Y.}~\bibnamefont
  {Dong}}, \bibinfo {author} {\bibfnamefont {B.}~\bibnamefont {Tian}}, \bibinfo
  {author} {\bibfnamefont {T.~J.}\ \bibnamefont {Kempa}}, \ and\ \bibinfo
  {author} {\bibfnamefont {C.~M.}\ \bibnamefont {Lieber}},\ }\href@noop {}
  {\bibfield  {journal} {\bibinfo  {journal} {Nano Letters}\ }\textbf {\bibinfo
  {volume} {9}},\ \bibinfo {pages} {2183} (\bibinfo {year} {2009})}\BibitemShut
  {NoStop}%
\bibitem [{\citenamefont {Yuan}\ \emph {et~al.}(2015)\citenamefont {Yuan},
  \citenamefont {Caroff}, \citenamefont {Wang}, \citenamefont {Guo},
  \citenamefont {Wang}, \citenamefont {Jackson}, \citenamefont {Smith},
  \citenamefont {Tan},\ and\ \citenamefont {Jagadish}}]{Yuan15}%
  \BibitemOpen
  \bibfield  {author} {\bibinfo {author} {\bibfnamefont {X.}~\bibnamefont
  {Yuan}}, \bibinfo {author} {\bibfnamefont {P.}~\bibnamefont {Caroff}},
  \bibinfo {author} {\bibfnamefont {F.}~\bibnamefont {Wang}}, \bibinfo {author}
  {\bibfnamefont {Y.}~\bibnamefont {Guo}}, \bibinfo {author} {\bibfnamefont
  {Y.}~\bibnamefont {Wang}}, \bibinfo {author} {\bibfnamefont {H.~E.}\
  \bibnamefont {Jackson}}, \bibinfo {author} {\bibfnamefont {L.~M.}\
  \bibnamefont {Smith}}, \bibinfo {author} {\bibfnamefont {H.~H.}\ \bibnamefont
  {Tan}}, \ and\ \bibinfo {author} {\bibfnamefont {C.}~\bibnamefont
  {Jagadish}},\ }\href@noop {} {\bibfield  {journal} {\bibinfo  {journal} {Adv.
  Funct. Mater.}\ }\textbf {\bibinfo {volume} {25}},\ \bibinfo {pages} {5300}
  (\bibinfo {year} {2015})}\BibitemShut {NoStop}%
\bibitem [{\citenamefont {G\"{o}ransson}\ \emph {et~al.}(2019)\citenamefont
  {G\"{o}ransson}, \citenamefont {Heurlin}, \citenamefont {Dalelkhan},
  \citenamefont {Abay}, \citenamefont {Messing}, \citenamefont {Maisi},
  \citenamefont {Borgström},\ and\ \citenamefont {Xu}}]{Goransson19}%
  \BibitemOpen
  \bibfield  {author} {\bibinfo {author} {\bibfnamefont {D.~J.~O.}\
  \bibnamefont {G\"{o}ransson}}, \bibinfo {author} {\bibfnamefont
  {M.}~\bibnamefont {Heurlin}}, \bibinfo {author} {\bibfnamefont
  {B.}~\bibnamefont {Dalelkhan}}, \bibinfo {author} {\bibfnamefont
  {S.}~\bibnamefont {Abay}}, \bibinfo {author} {\bibfnamefont {M.~E.}\
  \bibnamefont {Messing}}, \bibinfo {author} {\bibfnamefont {V.~F.}\
  \bibnamefont {Maisi}}, \bibinfo {author} {\bibfnamefont {M.~T.}\ \bibnamefont
  {Borgström}}, \ and\ \bibinfo {author} {\bibfnamefont {H.~Q.}\ \bibnamefont
  {Xu}},\ }\href {\doibase 10.1063/1.5084222} {\bibfield  {journal} {\bibinfo
  {journal} {Applied Physics Letters}\ }\textbf {\bibinfo {volume} {114}},\
  \bibinfo {pages} {053108} (\bibinfo {year} {2019})}\BibitemShut {NoStop}%
\bibitem [{\citenamefont {Rieger}\ \emph {et~al.}(2015)\citenamefont {Rieger},
  \citenamefont {Grutzmacher},\ and\ \citenamefont {Lepsa}}]{Rieger15}%
  \BibitemOpen
  \bibfield  {author} {\bibinfo {author} {\bibfnamefont {T.}~\bibnamefont
  {Rieger}}, \bibinfo {author} {\bibfnamefont {D.}~\bibnamefont {Grutzmacher}},
  \ and\ \bibinfo {author} {\bibfnamefont {M.~I.}\ \bibnamefont {Lepsa}},\
  }\href@noop {} {\bibfield  {journal} {\bibinfo  {journal} {Nanoscale}\
  }\textbf {\bibinfo {volume} {7}},\ \bibinfo {pages} {356} (\bibinfo {year}
  {2015})}\BibitemShut {NoStop}%
\bibitem [{\citenamefont {Fickenscher}\ \emph {et~al.}(2013)\citenamefont
  {Fickenscher}, \citenamefont {Shi}, \citenamefont {Jackson}, \citenamefont
  {Smith}, \citenamefont {Yarrison-Rice}, \citenamefont {Zheng}, \citenamefont
  {Miller}, \citenamefont {Etheridge}, \citenamefont {Wong}, \citenamefont
  {Gao}, \citenamefont {Deshpande}, \citenamefont {Tan},\ and\ \citenamefont
  {Jagadish}}]{Fickenscher13}%
  \BibitemOpen
  \bibfield  {author} {\bibinfo {author} {\bibfnamefont {M.}~\bibnamefont
  {Fickenscher}}, \bibinfo {author} {\bibfnamefont {T.}~\bibnamefont {Shi}},
  \bibinfo {author} {\bibfnamefont {H.~E.}\ \bibnamefont {Jackson}}, \bibinfo
  {author} {\bibfnamefont {L.~M.}\ \bibnamefont {Smith}}, \bibinfo {author}
  {\bibfnamefont {J.~M.}\ \bibnamefont {Yarrison-Rice}}, \bibinfo {author}
  {\bibfnamefont {C.}~\bibnamefont {Zheng}}, \bibinfo {author} {\bibfnamefont
  {P.}~\bibnamefont {Miller}}, \bibinfo {author} {\bibfnamefont
  {J.}~\bibnamefont {Etheridge}}, \bibinfo {author} {\bibfnamefont {B.~M.}\
  \bibnamefont {Wong}}, \bibinfo {author} {\bibfnamefont {Q.}~\bibnamefont
  {Gao}}, \bibinfo {author} {\bibfnamefont {S.}~\bibnamefont {Deshpande}},
  \bibinfo {author} {\bibfnamefont {H.~H.}\ \bibnamefont {Tan}}, \ and\
  \bibinfo {author} {\bibfnamefont {C.}~\bibnamefont {Jagadish}},\ }\href
  {\doibase 10.1021/nl304182j} {\bibfield  {journal} {\bibinfo  {journal} {Nano
  Letters}\ }\textbf {\bibinfo {volume} {13}},\ \bibinfo {pages} {1016}
  (\bibinfo {year} {2013})}\BibitemShut {NoStop}%
\bibitem [{\citenamefont {Shi}\ \emph {et~al.}(2015)\citenamefont {Shi},
  \citenamefont {Jackson}, \citenamefont {Smith}, \citenamefont {Jiang},
  \citenamefont {Gao}, \citenamefont {Tan}, \citenamefont {Jagadish},
  \citenamefont {Zheng},\ and\ \citenamefont {Etheridge}}]{Shi15}%
  \BibitemOpen
  \bibfield  {author} {\bibinfo {author} {\bibfnamefont {T.}~\bibnamefont
  {Shi}}, \bibinfo {author} {\bibfnamefont {H.~E.}\ \bibnamefont {Jackson}},
  \bibinfo {author} {\bibfnamefont {L.~M.}\ \bibnamefont {Smith}}, \bibinfo
  {author} {\bibfnamefont {N.}~\bibnamefont {Jiang}}, \bibinfo {author}
  {\bibfnamefont {Q.}~\bibnamefont {Gao}}, \bibinfo {author} {\bibfnamefont
  {H.~H.}\ \bibnamefont {Tan}}, \bibinfo {author} {\bibfnamefont
  {C.}~\bibnamefont {Jagadish}}, \bibinfo {author} {\bibfnamefont
  {C.}~\bibnamefont {Zheng}}, \ and\ \bibinfo {author} {\bibfnamefont
  {J.}~\bibnamefont {Etheridge}},\ }\href {\doibase 10.1021/nl5046878}
  {\bibfield  {journal} {\bibinfo  {journal} {Nano Letters}\ }\textbf {\bibinfo
  {volume} {15}},\ \bibinfo {pages} {1876} (\bibinfo {year}
  {2015})}\BibitemShut {NoStop}%
\bibitem [{\citenamefont {Kinzel}\ \emph {et~al.}(2016)\citenamefont {Kinzel},
  \citenamefont {Sch{\"u}lein}, \citenamefont {Wei\ss}, \citenamefont {Janker},
  \citenamefont {B{\"u}hler}, \citenamefont {Heigl}, \citenamefont {Rudolph},
  \citenamefont {Mork{\"o}tter}, \citenamefont {D{\"o}blinger}, \citenamefont
  {Bichler}, \citenamefont {Abstreiter}, \citenamefont {Finley}, \citenamefont
  {Wixforth}, \citenamefont {Koblm{\"u}ller},\ and\ \citenamefont
  {Krenner}}]{Kinzel16}%
  \BibitemOpen
  \bibfield  {author} {\bibinfo {author} {\bibfnamefont {J.~B.}\ \bibnamefont
  {Kinzel}}, \bibinfo {author} {\bibfnamefont {F.~J.~R.}\ \bibnamefont
  {Sch{\"u}lein}}, \bibinfo {author} {\bibfnamefont {M.}~\bibnamefont
  {Wei\ss}}, \bibinfo {author} {\bibfnamefont {L.}~\bibnamefont {Janker}},
  \bibinfo {author} {\bibfnamefont {D.~D.}\ \bibnamefont {B{\"u}hler}},
  \bibinfo {author} {\bibfnamefont {M.}~\bibnamefont {Heigl}}, \bibinfo
  {author} {\bibfnamefont {D.}~\bibnamefont {Rudolph}}, \bibinfo {author}
  {\bibfnamefont {S.}~\bibnamefont {Mork{\"o}tter}}, \bibinfo {author}
  {\bibfnamefont {M.}~\bibnamefont {D{\"o}blinger}}, \bibinfo {author}
  {\bibfnamefont {M.}~\bibnamefont {Bichler}}, \bibinfo {author} {\bibfnamefont
  {G.}~\bibnamefont {Abstreiter}}, \bibinfo {author} {\bibfnamefont {J.~J.}\
  \bibnamefont {Finley}}, \bibinfo {author} {\bibfnamefont {A.}~\bibnamefont
  {Wixforth}}, \bibinfo {author} {\bibfnamefont {G.}~\bibnamefont
  {Koblm{\"u}ller}}, \ and\ \bibinfo {author} {\bibfnamefont {H.~J.}\
  \bibnamefont {Krenner}},\ }\href@noop {} {\bibfield  {journal} {\bibinfo
  {journal} {ACS Nano}\ }\textbf {\bibinfo {volume} {10}},\ \bibinfo {pages}
  {4942} (\bibinfo {year} {2016})}\BibitemShut {NoStop}%
\bibitem [{\citenamefont {Li}\ \emph {et~al.}(2017)\citenamefont {Li},
  \citenamefont {Li}, \citenamefont {Tan}, \citenamefont {Zhou}, \citenamefont
  {Ma}, \citenamefont {Lysevych}, \citenamefont {Fu}, \citenamefont {Tan},\
  and\ \citenamefont {Jagadish}}]{Li17v2}%
  \BibitemOpen
  \bibfield  {author} {\bibinfo {author} {\bibfnamefont {F.}~\bibnamefont
  {Li}}, \bibinfo {author} {\bibfnamefont {Z.}~\bibnamefont {Li}}, \bibinfo
  {author} {\bibfnamefont {L.}~\bibnamefont {Tan}}, \bibinfo {author}
  {\bibfnamefont {Y.}~\bibnamefont {Zhou}}, \bibinfo {author} {\bibfnamefont
  {J.}~\bibnamefont {Ma}}, \bibinfo {author} {\bibfnamefont {M.}~\bibnamefont
  {Lysevych}}, \bibinfo {author} {\bibfnamefont {L.}~\bibnamefont {Fu}},
  \bibinfo {author} {\bibfnamefont {H.~H.}\ \bibnamefont {Tan}}, \ and\
  \bibinfo {author} {\bibfnamefont {C.}~\bibnamefont {Jagadish}},\ }\href@noop
  {} {\bibfield  {journal} {\bibinfo  {journal} {Nanotechnology}\ }\textbf
  {\bibinfo {volume} {28}},\ \bibinfo {pages} {125702} (\bibinfo {year}
  {2017})}\BibitemShut {NoStop}%
\bibitem [{\citenamefont {De~Luca}\ \emph {et~al.}(2013)\citenamefont
  {De~Luca}, \citenamefont {Lavenuta}, \citenamefont {Polimeni}, \citenamefont
  {Rubini}, \citenamefont {Grillo}, \citenamefont {Mura}, \citenamefont
  {Miriametro}, \citenamefont {Capizzi},\ and\ \citenamefont
  {Martelli}}]{DeLuca13}%
  \BibitemOpen
  \bibfield  {author} {\bibinfo {author} {\bibfnamefont {M.}~\bibnamefont
  {De~Luca}}, \bibinfo {author} {\bibfnamefont {G.}~\bibnamefont {Lavenuta}},
  \bibinfo {author} {\bibfnamefont {A.}~\bibnamefont {Polimeni}}, \bibinfo
  {author} {\bibfnamefont {S.}~\bibnamefont {Rubini}}, \bibinfo {author}
  {\bibfnamefont {V.}~\bibnamefont {Grillo}}, \bibinfo {author} {\bibfnamefont
  {F.}~\bibnamefont {Mura}}, \bibinfo {author} {\bibfnamefont {A.}~\bibnamefont
  {Miriametro}}, \bibinfo {author} {\bibfnamefont {M.}~\bibnamefont {Capizzi}},
  \ and\ \bibinfo {author} {\bibfnamefont {F.}~\bibnamefont {Martelli}},\
  }\href@noop {} {\bibfield  {journal} {\bibinfo  {journal} {Phys. Rev. B}\
  }\textbf {\bibinfo {volume} {87}},\ \bibinfo {pages} {235304} (\bibinfo
  {year} {2013})}\BibitemShut {NoStop}%
\bibitem [{\citenamefont {Plochocka}\ \emph {et~al.}(2013)\citenamefont
  {Plochocka}, \citenamefont {Mitioglu}, \citenamefont {Maude}, \citenamefont
  {Rikken}, \citenamefont {Granados~del Águila}, \citenamefont {Christianen},
  \citenamefont {Kacman},\ and\ \citenamefont {Shtrikman}}]{Plochocka13}%
  \BibitemOpen
  \bibfield  {author} {\bibinfo {author} {\bibfnamefont {P.}~\bibnamefont
  {Plochocka}}, \bibinfo {author} {\bibfnamefont {A.~A.}\ \bibnamefont
  {Mitioglu}}, \bibinfo {author} {\bibfnamefont {D.~K.}\ \bibnamefont {Maude}},
  \bibinfo {author} {\bibfnamefont {G.~L. J.~A.}\ \bibnamefont {Rikken}},
  \bibinfo {author} {\bibfnamefont {A.}~\bibnamefont {Granados~del Águila}},
  \bibinfo {author} {\bibfnamefont {P.~C.~M.}\ \bibnamefont {Christianen}},
  \bibinfo {author} {\bibfnamefont {P.}~\bibnamefont {Kacman}}, \ and\ \bibinfo
  {author} {\bibfnamefont {H.}~\bibnamefont {Shtrikman}},\ }\href@noop {}
  {\bibfield  {journal} {\bibinfo  {journal} {Nano Letters}\ }\textbf {\bibinfo
  {volume} {13}},\ \bibinfo {pages} {2442} (\bibinfo {year}
  {2013})}\BibitemShut {NoStop}%
\bibitem [{\citenamefont {Buscemi}\ \emph {et~al.}(2015)\citenamefont
  {Buscemi}, \citenamefont {Royo}, \citenamefont {Bertoni},\ and\ \citenamefont
  {Goldoni}}]{Buscemi15}%
  \BibitemOpen
  \bibfield  {author} {\bibinfo {author} {\bibfnamefont {F.}~\bibnamefont
  {Buscemi}}, \bibinfo {author} {\bibfnamefont {M.}~\bibnamefont {Royo}},
  \bibinfo {author} {\bibfnamefont {A.}~\bibnamefont {Bertoni}}, \ and\
  \bibinfo {author} {\bibfnamefont {G.}~\bibnamefont {Goldoni}},\ }\href@noop
  {} {\bibfield  {journal} {\bibinfo  {journal} {Phys. Rev. B}\ }\textbf
  {\bibinfo {volume} {92}},\ \bibinfo {pages} {165302} (\bibinfo {year}
  {2015})}\BibitemShut {NoStop}%
\bibitem [{\citenamefont {Ferrari}\ \emph {et~al.}(2009)\citenamefont
  {Ferrari}, \citenamefont {Goldoni}, \citenamefont {Bertoni}, \citenamefont
  {Cuoghi},\ and\ \citenamefont {Molinari}}]{Ferrari09b}%
  \BibitemOpen
  \bibfield  {author} {\bibinfo {author} {\bibfnamefont {G.}~\bibnamefont
  {Ferrari}}, \bibinfo {author} {\bibfnamefont {G.}~\bibnamefont {Goldoni}},
  \bibinfo {author} {\bibfnamefont {A.}~\bibnamefont {Bertoni}}, \bibinfo
  {author} {\bibfnamefont {G.}~\bibnamefont {Cuoghi}}, \ and\ \bibinfo {author}
  {\bibfnamefont {E.}~\bibnamefont {Molinari}},\ }\href@noop {} {\bibfield
  {journal} {\bibinfo  {journal} {Nano Letters}\ }\textbf {\bibinfo {volume}
  {9}},\ \bibinfo {pages} {1631} (\bibinfo {year} {2009})}\BibitemShut
  {NoStop}%
\bibitem [{\citenamefont {Wong}\ \emph {et~al.}(2011)\citenamefont {Wong},
  \citenamefont {Léonard}, \citenamefont {Li},\ and\ \citenamefont
  {Wang}}]{Wong11}%
  \BibitemOpen
  \bibfield  {author} {\bibinfo {author} {\bibfnamefont {B.~M.}\ \bibnamefont
  {Wong}}, \bibinfo {author} {\bibfnamefont {F.}~\bibnamefont {Léonard}},
  \bibinfo {author} {\bibfnamefont {Q.}~\bibnamefont {Li}}, \ and\ \bibinfo
  {author} {\bibfnamefont {G.~T.}\ \bibnamefont {Wang}},\ }\href@noop {}
  {\bibfield  {journal} {\bibinfo  {journal} {Nano Letters}\ }\textbf {\bibinfo
  {volume} {11}},\ \bibinfo {pages} {3074} (\bibinfo {year}
  {2011})}\BibitemShut {NoStop}%
\bibitem [{\citenamefont {Li}\ \emph {et~al.}(2018)\citenamefont {Li},
  \citenamefont {Alradhi}, \citenamefont {Jin}, \citenamefont {Anyebe},
  \citenamefont {Sanchez}, \citenamefont {Linhart}, \citenamefont {Kudrawiec},
  \citenamefont {Fang}, \citenamefont {Wang}, \citenamefont {Hu},\ and\
  \citenamefont {Zhuang}}]{Li18}%
  \BibitemOpen
  \bibfield  {author} {\bibinfo {author} {\bibfnamefont {H.}~\bibnamefont
  {Li}}, \bibinfo {author} {\bibfnamefont {H.}~\bibnamefont {Alradhi}},
  \bibinfo {author} {\bibfnamefont {Z.}~\bibnamefont {Jin}}, \bibinfo {author}
  {\bibfnamefont {E.~A.}\ \bibnamefont {Anyebe}}, \bibinfo {author}
  {\bibfnamefont {A.~M.}\ \bibnamefont {Sanchez}}, \bibinfo {author}
  {\bibfnamefont {W.~M.}\ \bibnamefont {Linhart}}, \bibinfo {author}
  {\bibfnamefont {R.}~\bibnamefont {Kudrawiec}}, \bibinfo {author}
  {\bibfnamefont {H.}~\bibnamefont {Fang}}, \bibinfo {author} {\bibfnamefont
  {Z.}~\bibnamefont {Wang}}, \bibinfo {author} {\bibfnamefont {W.}~\bibnamefont
  {Hu}}, \ and\ \bibinfo {author} {\bibfnamefont {Q.}~\bibnamefont {Zhuang}},\
  }\href@noop {} {\bibfield  {journal} {\bibinfo  {journal} {Advanced
  Functional Materials}\ }\textbf {\bibinfo {volume} {28}},\ \bibinfo {pages}
  {1705382} (\bibinfo {year} {2018})}\BibitemShut {NoStop}%
\bibitem [{\citenamefont {Sitek}\ \emph
  {et~al.}(2015{\natexlab{a}})\citenamefont {Sitek}, \citenamefont {Serra},
  \citenamefont {Gudmundsson},\ and\ \citenamefont {Manolescu}}]{Sitek15}%
  \BibitemOpen
  \bibfield  {author} {\bibinfo {author} {\bibfnamefont {A.}~\bibnamefont
  {Sitek}}, \bibinfo {author} {\bibfnamefont {L.}~\bibnamefont {Serra}},
  \bibinfo {author} {\bibfnamefont {V.}~\bibnamefont {Gudmundsson}}, \ and\
  \bibinfo {author} {\bibfnamefont {A.}~\bibnamefont {Manolescu}},\ }\href@noop
  {} {\bibfield  {journal} {\bibinfo  {journal} {Phys. Rev. B}\ }\textbf
  {\bibinfo {volume} {91}},\ \bibinfo {pages} {235429} (\bibinfo {year}
  {2015}{\natexlab{a}})}\BibitemShut {NoStop}%
\bibitem [{\citenamefont {Sitek}\ \emph {et~al.}(2016)\citenamefont {Sitek},
  \citenamefont {Thorgilsson}, \citenamefont {Gudmundsson},\ and\ \citenamefont
  {Manolescu}}]{Sitek16}%
  \BibitemOpen
  \bibfield  {author} {\bibinfo {author} {\bibfnamefont {A.}~\bibnamefont
  {Sitek}}, \bibinfo {author} {\bibfnamefont {G.}~\bibnamefont {Thorgilsson}},
  \bibinfo {author} {\bibfnamefont {V.}~\bibnamefont {Gudmundsson}}, \ and\
  \bibinfo {author} {\bibfnamefont {A.}~\bibnamefont {Manolescu}},\ }\href@noop
  {} {\bibfield  {journal} {\bibinfo  {journal} {Nanotechnology}\ }\textbf
  {\bibinfo {volume} {27}},\ \bibinfo {pages} {225202} (\bibinfo {year}
  {2016})}\BibitemShut {NoStop}%
\bibitem [{\citenamefont {Sitek}\ \emph {et~al.}(2018)\citenamefont {Sitek},
  \citenamefont {Urbaneja~Torres}, \citenamefont {Torfason}, \citenamefont
  {Gudmundsson}, \citenamefont {Bertoni},\ and\ \citenamefont
  {Manolescu}}]{Sitek18}%
  \BibitemOpen
  \bibfield  {author} {\bibinfo {author} {\bibfnamefont {A.}~\bibnamefont
  {Sitek}}, \bibinfo {author} {\bibfnamefont {M.}~\bibnamefont
  {Urbaneja~Torres}}, \bibinfo {author} {\bibfnamefont {K.}~\bibnamefont
  {Torfason}}, \bibinfo {author} {\bibfnamefont {V.}~\bibnamefont
  {Gudmundsson}}, \bibinfo {author} {\bibfnamefont {A.}~\bibnamefont
  {Bertoni}}, \ and\ \bibinfo {author} {\bibfnamefont {A.}~\bibnamefont
  {Manolescu}},\ }\href@noop {} {\bibfield  {journal} {\bibinfo  {journal}
  {Nano Letters}\ }\textbf {\bibinfo {volume} {18}},\ \bibinfo {pages} {2581}
  (\bibinfo {year} {2018})}\BibitemShut {NoStop}%
\bibitem [{\citenamefont {Urbaneja~Torres}\ \emph {et~al.}(2018)\citenamefont
  {Urbaneja~Torres}, \citenamefont {Sitek}, \citenamefont {Erlingsson},
  \citenamefont {Thorgilsson}, \citenamefont {Gudmundsson},\ and\ \citenamefont
  {Manolescu}}]{Urbaneja18}%
  \BibitemOpen
  \bibfield  {author} {\bibinfo {author} {\bibfnamefont {M.}~\bibnamefont
  {Urbaneja~Torres}}, \bibinfo {author} {\bibfnamefont {A.}~\bibnamefont
  {Sitek}}, \bibinfo {author} {\bibfnamefont {S.~I.}\ \bibnamefont
  {Erlingsson}}, \bibinfo {author} {\bibfnamefont {G.}~\bibnamefont
  {Thorgilsson}}, \bibinfo {author} {\bibfnamefont {V.}~\bibnamefont
  {Gudmundsson}}, \ and\ \bibinfo {author} {\bibfnamefont {A.}~\bibnamefont
  {Manolescu}},\ }\href@noop {} {\bibfield  {journal} {\bibinfo  {journal}
  {Phys. Rev. B}\ }\textbf {\bibinfo {volume} {98}},\ \bibinfo {pages} {085419}
  (\bibinfo {year} {2018})}\BibitemShut {NoStop}%
\bibitem [{\citenamefont {Manolescu}\ \emph {et~al.}(2017)\citenamefont
  {Manolescu}, \citenamefont {Sitek}, \citenamefont {Osca}, \citenamefont
  {Serra}, \citenamefont {Gudmundsson},\ and\ \citenamefont
  {Stanescu}}]{Manolescu17}%
  \BibitemOpen
  \bibfield  {author} {\bibinfo {author} {\bibfnamefont {A.}~\bibnamefont
  {Manolescu}}, \bibinfo {author} {\bibfnamefont {A.}~\bibnamefont {Sitek}},
  \bibinfo {author} {\bibfnamefont {J.}~\bibnamefont {Osca}}, \bibinfo {author}
  {\bibfnamefont {L.}~\bibnamefont {Serra}}, \bibinfo {author} {\bibfnamefont
  {V.}~\bibnamefont {Gudmundsson}}, \ and\ \bibinfo {author} {\bibfnamefont
  {T.~D.}\ \bibnamefont {Stanescu}},\ }\href@noop {} {\bibfield  {journal}
  {\bibinfo  {journal} {Phys. Rev. B}\ }\textbf {\bibinfo {volume} {96}},\
  \bibinfo {pages} {125435} (\bibinfo {year} {2017})}\BibitemShut {NoStop}%
\bibitem [{\citenamefont {Stanescu}\ \emph {et~al.}(2018)\citenamefont
  {Stanescu}, \citenamefont {Sitek},\ and\ \citenamefont
  {Manolescu}}]{Stanescu18}%
  \BibitemOpen
  \bibfield  {author} {\bibinfo {author} {\bibfnamefont {T.~D.}\ \bibnamefont
  {Stanescu}}, \bibinfo {author} {\bibfnamefont {A.}~\bibnamefont {Sitek}}, \
  and\ \bibinfo {author} {\bibfnamefont {A.}~\bibnamefont {Manolescu}},\
  }\href@noop {} {\bibfield  {journal} {\bibinfo  {journal} {Beilstein J.
  Nanotechnol.}\ }\textbf {\bibinfo {volume} {9}},\ \bibinfo {pages} {1512}
  (\bibinfo {year} {2018})}\BibitemShut {NoStop}%
\bibitem [{\citenamefont {Sitek}\ \emph {et~al.}(2017)\citenamefont {Sitek},
  \citenamefont {\c{T}olea}, \citenamefont {Ni\c{t}\u{a}}, \citenamefont
  {Serra}, \citenamefont {Gudmundsson},\ and\ \citenamefont
  {Manolescu}}]{Sitek17}%
  \BibitemOpen
  \bibfield  {author} {\bibinfo {author} {\bibfnamefont {A.}~\bibnamefont
  {Sitek}}, \bibinfo {author} {\bibfnamefont {M.}~\bibnamefont {\c{T}olea}},
  \bibinfo {author} {\bibfnamefont {M.}~\bibnamefont {Ni\c{t}\u{a}}}, \bibinfo
  {author} {\bibfnamefont {L.}~\bibnamefont {Serra}}, \bibinfo {author}
  {\bibfnamefont {V.}~\bibnamefont {Gudmundsson}}, \ and\ \bibinfo {author}
  {\bibfnamefont {A.}~\bibnamefont {Manolescu}},\ }\href@noop {} {\bibfield
  {journal} {\bibinfo  {journal} {Sci. Rep.}\ }\textbf {\bibinfo {volume}
  {7}},\ \bibinfo {pages} {40197} (\bibinfo {year} {2017})}\BibitemShut
  {NoStop}%
\bibitem [{\citenamefont {G\"ul}\ \emph {et~al.}(2014)\citenamefont {G\"ul},
  \citenamefont {Demarina}, \citenamefont {Bl\"omers}, \citenamefont {Rieger},
  \citenamefont {L\"uth}, \citenamefont {Lepsa}, \citenamefont
  {Gr\"utzmacher},\ and\ \citenamefont {Sch\"apers}}]{Gul14}%
  \BibitemOpen
  \bibfield  {author} {\bibinfo {author} {\bibfnamefont {O.}~\bibnamefont
  {G\"ul}}, \bibinfo {author} {\bibfnamefont {N.}~\bibnamefont {Demarina}},
  \bibinfo {author} {\bibfnamefont {C.}~\bibnamefont {Bl\"omers}}, \bibinfo
  {author} {\bibfnamefont {T.}~\bibnamefont {Rieger}}, \bibinfo {author}
  {\bibfnamefont {H.}~\bibnamefont {L\"uth}}, \bibinfo {author} {\bibfnamefont
  {M.~I.}\ \bibnamefont {Lepsa}}, \bibinfo {author} {\bibfnamefont
  {D.}~\bibnamefont {Gr\"utzmacher}}, \ and\ \bibinfo {author} {\bibfnamefont
  {T.}~\bibnamefont {Sch\"apers}},\ }\href@noop {} {\bibfield  {journal}
  {\bibinfo  {journal} {Phys. Rev. B}\ }\textbf {\bibinfo {volume} {89}},\
  \bibinfo {pages} {045417} (\bibinfo {year} {2014})}\BibitemShut {NoStop}%
\bibitem [{\citenamefont {Rosdahl}\ \emph {et~al.}(2014)\citenamefont
  {Rosdahl}, \citenamefont {Manolescu},\ and\ \citenamefont
  {Gudmundsson}}]{Rosdahl14}%
  \BibitemOpen
  \bibfield  {author} {\bibinfo {author} {\bibfnamefont {T.~O.}\ \bibnamefont
  {Rosdahl}}, \bibinfo {author} {\bibfnamefont {A.}~\bibnamefont {Manolescu}},
  \ and\ \bibinfo {author} {\bibfnamefont {V.}~\bibnamefont {Gudmundsson}},\
  }\href@noop {} {\bibfield  {journal} {\bibinfo  {journal} {Phys. Rev. B}\
  }\textbf {\bibinfo {volume} {90}},\ \bibinfo {pages} {035421} (\bibinfo
  {year} {2014})}\BibitemShut {NoStop}%
\bibitem [{\citenamefont {Heedt}\ \emph {et~al.}(2016)\citenamefont {Heedt},
  \citenamefont {Manolescu}, \citenamefont {Nemnes}, \citenamefont {Prost},
  \citenamefont {Schubert}, \citenamefont {Gr\"utzmacher},\ and\ \citenamefont
  {Sch\"apers}}]{Heedt16}%
  \BibitemOpen
  \bibfield  {author} {\bibinfo {author} {\bibfnamefont {S.}~\bibnamefont
  {Heedt}}, \bibinfo {author} {\bibfnamefont {A.}~\bibnamefont {Manolescu}},
  \bibinfo {author} {\bibfnamefont {G.~A.}\ \bibnamefont {Nemnes}}, \bibinfo
  {author} {\bibfnamefont {W.}~\bibnamefont {Prost}}, \bibinfo {author}
  {\bibfnamefont {J.}~\bibnamefont {Schubert}}, \bibinfo {author}
  {\bibfnamefont {D.}~\bibnamefont {Gr\"utzmacher}}, \ and\ \bibinfo {author}
  {\bibfnamefont {T.}~\bibnamefont {Sch\"apers}},\ }\href {\doibase
  10.1021/acs.nanolett.6b01840} {\bibfield  {journal} {\bibinfo  {journal}
  {Nano Letters}\ }\textbf {\bibinfo {volume} {16}},\ \bibinfo {pages} {4569}
  (\bibinfo {year} {2016})}\BibitemShut {NoStop}%
\bibitem [{\citenamefont {Battiato}\ \emph {et~al.}(2019)\citenamefont
  {Battiato}, \citenamefont {Wu}, \citenamefont {Zannier}, \citenamefont
  {Bertoni}, \citenamefont {Goldoni}, \citenamefont {Li}, \citenamefont {Xiao},
  \citenamefont {Han}, \citenamefont {Beltram}, \citenamefont {Sorba},
  \citenamefont {Xu},\ and\ \citenamefont {Rossella}}]{Battiato19}%
  \BibitemOpen
  \bibfield  {author} {\bibinfo {author} {\bibfnamefont {S.}~\bibnamefont
  {Battiato}}, \bibinfo {author} {\bibfnamefont {S.}~\bibnamefont {Wu}},
  \bibinfo {author} {\bibfnamefont {V.}~\bibnamefont {Zannier}}, \bibinfo
  {author} {\bibfnamefont {A.}~\bibnamefont {Bertoni}}, \bibinfo {author}
  {\bibfnamefont {G.}~\bibnamefont {Goldoni}}, \bibinfo {author} {\bibfnamefont
  {A.}~\bibnamefont {Li}}, \bibinfo {author} {\bibfnamefont {S.}~\bibnamefont
  {Xiao}}, \bibinfo {author} {\bibfnamefont {X.~D.}\ \bibnamefont {Han}},
  \bibinfo {author} {\bibfnamefont {F.}~\bibnamefont {Beltram}}, \bibinfo
  {author} {\bibfnamefont {L.}~\bibnamefont {Sorba}}, \bibinfo {author}
  {\bibfnamefont {X.}~\bibnamefont {Xu}}, \ and\ \bibinfo {author}
  {\bibfnamefont {F.}~\bibnamefont {Rossella}},\ }\href {\doibase
  10.1088/1361-6528/aafde4} {\bibfield  {journal} {\bibinfo  {journal}
  {Nanotechnology}\ }\textbf {\bibinfo {volume} {30}},\ \bibinfo {pages}
  {194004} (\bibinfo {year} {2019})}\BibitemShut {NoStop}%
\bibitem [{\citenamefont {Sonner}\ \emph {et~al.}(2019)\citenamefont {Sonner},
  \citenamefont {Sitek}, \citenamefont {Janker}, \citenamefont {Rudolph},
  \citenamefont {Ruhstorfer}, \citenamefont {D{\"o}blinger}, \citenamefont
  {Manolescu}, \citenamefont {Abstreiter}, \citenamefont {Finley},
  \citenamefont {Wixforth}, \citenamefont {Koblmüller},\ and\ \citenamefont
  {Krenner}}]{Sonner19}%
  \BibitemOpen
  \bibfield  {author} {\bibinfo {author} {\bibfnamefont {M.~M.}\ \bibnamefont
  {Sonner}}, \bibinfo {author} {\bibfnamefont {A.}~\bibnamefont {Sitek}},
  \bibinfo {author} {\bibfnamefont {L.}~\bibnamefont {Janker}}, \bibinfo
  {author} {\bibfnamefont {D.}~\bibnamefont {Rudolph}}, \bibinfo {author}
  {\bibfnamefont {D.}~\bibnamefont {Ruhstorfer}}, \bibinfo {author}
  {\bibfnamefont {M.}~\bibnamefont {D{\"o}blinger}}, \bibinfo {author}
  {\bibfnamefont {A.}~\bibnamefont {Manolescu}}, \bibinfo {author}
  {\bibfnamefont {G.}~\bibnamefont {Abstreiter}}, \bibinfo {author}
  {\bibfnamefont {J.~J.}\ \bibnamefont {Finley}}, \bibinfo {author}
  {\bibfnamefont {A.}~\bibnamefont {Wixforth}}, \bibinfo {author}
  {\bibfnamefont {G.}~\bibnamefont {Koblmüller}}, \ and\ \bibinfo {author}
  {\bibfnamefont {H.~J.}\ \bibnamefont {Krenner}},\ }\href@noop {} {\bibfield
  {journal} {\bibinfo  {journal} {Nano Letters}\ }\textbf {\bibinfo {volume}
  {19}},\ \bibinfo {pages} {3336} (\bibinfo {year} {2019})}\BibitemShut
  {NoStop}%
\bibitem [{\citenamefont {Daday}\ \emph {et~al.}(2011)\citenamefont {Daday},
  \citenamefont {Manolescu}, \citenamefont {Marinescu},\ and\ \citenamefont
  {Gudmundsson}}]{Daday11}%
  \BibitemOpen
  \bibfield  {author} {\bibinfo {author} {\bibfnamefont {C.}~\bibnamefont
  {Daday}}, \bibinfo {author} {\bibfnamefont {A.}~\bibnamefont {Manolescu}},
  \bibinfo {author} {\bibfnamefont {D.~C.}\ \bibnamefont {Marinescu}}, \ and\
  \bibinfo {author} {\bibfnamefont {V.}~\bibnamefont {Gudmundsson}},\
  }\href@noop {} {\bibfield  {journal} {\bibinfo  {journal} {Phys. Rev. B}\
  }\textbf {\bibinfo {volume} {84}},\ \bibinfo {pages} {115311} (\bibinfo
  {year} {2011})}\BibitemShut {NoStop}%
\bibitem [{\citenamefont {Ballester}\ \emph {et~al.}(2012)\citenamefont
  {Ballester}, \citenamefont {Planelles},\ and\ \citenamefont
  {Bertoni}}]{Ballester12}%
  \BibitemOpen
  \bibfield  {author} {\bibinfo {author} {\bibfnamefont {A.}~\bibnamefont
  {Ballester}}, \bibinfo {author} {\bibfnamefont {J.}~\bibnamefont
  {Planelles}}, \ and\ \bibinfo {author} {\bibfnamefont {A.}~\bibnamefont
  {Bertoni}},\ }\href@noop {} {\bibfield  {journal} {\bibinfo  {journal}
  {Journal of Applied Physics}\ }\textbf {\bibinfo {volume} {112}},\ \bibinfo
  {eid} {104317} (\bibinfo {year} {2012})}\BibitemShut {NoStop}%
\bibitem [{\citenamefont {G\"uniat}\ \emph {et~al.}(2019)\citenamefont
  {G\"uniat}, \citenamefont {Martí-Sánchez}, \citenamefont {Garcia},
  \citenamefont {Boscardin}, \citenamefont {Vindice}, \citenamefont {Tappy},
  \citenamefont {Friedl}, \citenamefont {Kim}, \citenamefont {Zamani},
  \citenamefont {Francaviglia}, \citenamefont {Balgarkashi}, \citenamefont
  {Leran}, \citenamefont {Arbiol},\ and\ \citenamefont {Fontcuberta~i
  Morral}}]{Guniat19}%
  \BibitemOpen
  \bibfield  {author} {\bibinfo {author} {\bibfnamefont {L.}~\bibnamefont
  {G\"uniat}}, \bibinfo {author} {\bibfnamefont {S.}~\bibnamefont
  {Martí-Sánchez}}, \bibinfo {author} {\bibfnamefont {O.}~\bibnamefont
  {Garcia}}, \bibinfo {author} {\bibfnamefont {M.}~\bibnamefont {Boscardin}},
  \bibinfo {author} {\bibfnamefont {D.}~\bibnamefont {Vindice}}, \bibinfo
  {author} {\bibfnamefont {N.}~\bibnamefont {Tappy}}, \bibinfo {author}
  {\bibfnamefont {M.}~\bibnamefont {Friedl}}, \bibinfo {author} {\bibfnamefont
  {W.}~\bibnamefont {Kim}}, \bibinfo {author} {\bibfnamefont {M.}~\bibnamefont
  {Zamani}}, \bibinfo {author} {\bibfnamefont {L.}~\bibnamefont
  {Francaviglia}}, \bibinfo {author} {\bibfnamefont {A.}~\bibnamefont
  {Balgarkashi}}, \bibinfo {author} {\bibfnamefont {J.-B.}\ \bibnamefont
  {Leran}}, \bibinfo {author} {\bibfnamefont {J.}~\bibnamefont {Arbiol}}, \
  and\ \bibinfo {author} {\bibfnamefont {A.}~\bibnamefont {Fontcuberta~i
  Morral}},\ }\href {\doibase 10.1021/acsnano.9b01546} {\bibfield  {journal}
  {\bibinfo  {journal} {ACS Nano}\ }\textbf {\bibinfo {volume} {13}},\ \bibinfo
  {pages} {5833} (\bibinfo {year} {2019})}\BibitemShut {NoStop}%
\bibitem [{\citenamefont {Wu}\ \emph {et~al.}(1992)\citenamefont {Wu},
  \citenamefont {Sprung},\ and\ \citenamefont {Martorell}}]{Wu}%
  \BibitemOpen
  \bibfield  {author} {\bibinfo {author} {\bibfnamefont {H.}~\bibnamefont
  {Wu}}, \bibinfo {author} {\bibfnamefont {D.~W.~L.}\ \bibnamefont {Sprung}}, \
  and\ \bibinfo {author} {\bibfnamefont {J.}~\bibnamefont {Martorell}},\ }\href
  {\doibase 10.1103/PhysRevB.45.11960} {\bibfield  {journal} {\bibinfo
  {journal} {Phys. Rev. B}\ }\textbf {\bibinfo {volume} {45}},\ \bibinfo
  {pages} {11960} (\bibinfo {year} {1992})}\BibitemShut {NoStop}%
\bibitem [{\citenamefont {Ameruddin}\ \emph {et~al.}(2015)\citenamefont
  {Ameruddin} \emph {et~al.}}]{Ameruddin15}%
  \BibitemOpen
  \bibfield  {author} {\bibinfo {author} {\bibfnamefont {A.~S.}\ \bibnamefont
  {Ameruddin}} \emph {et~al.},\ }\href@noop {} {\bibfield  {journal} {\bibinfo
  {journal} {Nanotechnology}\ }\textbf {\bibinfo {volume} {26}},\ \bibinfo
  {pages} {205604} (\bibinfo {year} {2015})}\BibitemShut {NoStop}%
\bibitem [{\citenamefont {Sitek}\ \emph
  {et~al.}(2015{\natexlab{b}})\citenamefont {Sitek}, \citenamefont
  {Gudmundsson},\ and\ \citenamefont {Manolescu}}]{Sitek15ICTON}%
  \BibitemOpen
  \bibfield  {author} {\bibinfo {author} {\bibfnamefont {A.}~\bibnamefont
  {Sitek}}, \bibinfo {author} {\bibfnamefont {V.}~\bibnamefont {Gudmundsson}},
  \ and\ \bibinfo {author} {\bibfnamefont {A.}~\bibnamefont {Manolescu}},\
  }\href {\doibase 10.1109/ICTON.2015.7193541} {\bibfield  {journal} {\bibinfo
  {journal} {Proceedings of the 17th International Conference on Transparent
  Optical Networks (ICTON 2015)}\ } (\bibinfo {year} {2015}{\natexlab{b}}),\
  10.1109/ICTON.2015.7193541}\BibitemShut {NoStop}%
\end{thebibliography}%

\end{document}